\documentclass[openacc]{rsproca_new}

\newcommand*\diff{\mathop{}\!\mathrm{d}}
\usepackage{bm}
\DeclareMathOperator{\tr}{tr}
\newcommand{\me}{\mathrm{e}}
\newcommand{\mi}{\mathrm{i}}
\usepackage{mathrsfs}
\usepackage{siunitx}
\usepackage{subcaption}
\newcommand\numberthis{\addtocounter{equation}{1}\tag{\theequation}}
\usepackage{diagbox}
 \usepackage[numbers]{natbib}

\begin{document}

\title{A unified framework for linear thermo-visco-elastic wave propagation including the effects of stress-relaxation}

\author{
Erik~Garc\'ia~Neefjes$^{1}$, David~Nigro$^{2}$, Artur~L.~Gower$^{3}$, Rapha\"{e}l~C.~Assier$^{1}$, Valerie J. Pinfield$^{4}$, William~J.~Parnell$^{1}$}

\address{$^{1}$Department of Mathematics, University of Manchester, Oxford Rd, Manchester, M13 9PL, UK\\
$^{2}$Thales United Kingdom, 350 Longwater Avenue Green Park, Reading RG2 6GF, UK\\
$^{3}$Department of Mechanical Engineering, University of Sheffield, Sheffield, UK \\
$^{4}$
Department of Chemical Engineering, Loughborough University, Loughborough, LE11 3TU, UK}

\subject{Acoustics, Applied Mathematics,
Mechanics, Wave Motion}

\keywords{Thermo-visco-elasticity, wave propagation, stress relaxation}

\corres{William J.\ Parnell\\
\email{william.parnell@manchester.ac.uk}}

\begin{abstract}

We present a unified framework for the study of wave propagation in homogeneous linear thermo-visco-elastic (TVE) continua, starting from conservation laws. In free-space such media admit two thermo-compressional modes and a shear mode. We provide asymptotic approximations to the corresponding wavenumbers which facilitate the understanding of dispersion of these modes, and consider common solids and fluids as well as soft materials where creep compliance and stress relaxation are important. 

We further illustrate how commonly used simpler acoustic/elastic dissipative theories can be derived via particular limits of this framework. Consequently, our framework allows us to: i) simultaneously model interfaces involving both fluids and solids, and ii) easily quantify the influence of thermal or viscous losses in a given configuration of interest.

As an example, the general framework is applied to the canonical problem of scattering from an interface between two TVE half spaces in perfect contact. To illustrate, we provide results for fluid-solid interfaces involving air, water, steel and rubber, paying particular attention to the effects of stress relaxation.
 
\end{abstract}
\maketitle

\section{Introduction}
Even under small deformations, complex continua exhibit a variety of constitutive effects over a broad  range of frequencies, associated with their atomistic, molecular or mesoscopic properties. In the field of continuum mechanics it has become common place to label a material either as  \textit{fluid} or \textit{solid} and even when viscoelastic, for reasons of model simplification, there is a tendency to specify a medium as a viscoelastic \textit{fluid} or viscoelastic \textit{solid}. This matter is however made more complex when considering wave propagation in the medium over a wide range of frequencies and temperatures. Polymers are an exemplary example; they are fluid-like at low frequencies and solid-like at high frequencies and they take on similar properties as a (reciprocal) function of temperature \cite{tobolsky1952elastoviscous,obaid2017understanding}. The shear modulus of a polymeric material can vary by several orders of magnitude after transitioning through the glass-transition frequency/temperature \cite{jeong1987frequency, kari2001dynamic}.

Assuming specific material behaviour can be helpful to limit the number of parameters that have to be measured experimentally, but it can also, unintentionally, lead to additional complexities. For example, consider two \textit{homogeneous} continua coupled at an interface, with the first being an \textit{acoustic} medium, such that only compressional (longitudinal) waves propagate, while the second is an \textit{elastic} medium which supports both shear (transverse) and compressional waves and displacement is vectorial \cite{cotterill2018thermo}. Combing the boundary conditions for acoustic and elastic waves can be awkward. This is specially true when it is important to include thermo-viscous effects loss~\cite{karlsen2015forces, cotterill2018thermo}. A unified framework, as we present here, simplifies the calculations as there is no need to develop separate models (i.e. one for the acoustic and the other for the elastic), and types of boundary conditions.

In the context of thermo-visco-acoustics, effective boundary conditions have been  devised to simplify the problem \cite{bruneau2013fundamentals} but when strong coupling occurs, this same approach cannot be employed. What follows are then rather ad-hoc approaches and often questionable approximations, particularly with regard to modelling in the time domain.

In more complex, inhomogeneous media or \textit{metamaterials}, the frequency dependence can often be very strong due to inherent resonances associated with microstructure \cite{cummer2016controlling, jimenez2017rainbow}. These resonances are often tuned to be strong at low frequencies, given that this is often the regime in which traditional materials cannot yield dispersive effects. However the frequency dependence is tuned by resonator size and geometry, and the material properties of the matrix medium.

Understanding the wave propagation characteristics of metamaterials is frequently achieved by employing asymptotic theories, which rely on specific scalings of the material property contrast \cite{pham2017two,touboul2020effective}. If this dependence changes with frequency then the entire theory underpinning these materials could be described as unstable. And the kinds of materials involved in high contrast resonance \textit{are} precisely materials that would possess strong frequency dependence \cite{krushynska2016visco}.  
 
One may argue that experiments at fixed frequencies can be fitted to a theory with certain fixed parameters, whether one considers a metamaterial or a simple, homogeneous medium. This is certainly the case and this approach has been employed very successfully in the past \cite{fernandez2019aerogel, favretto1997excitation}. However one may reasonably ask what happens when we change frequency, or design a resonator in the same matrix material to act at a different frequency, or even more reasonably what happens in the time domain? In all of these cases, of crucial importance is the ability to model the material's behaviour properly in the frequency domain. It appears uncomfortable from both a practical and scientific perspective to fit different parameters to the behaviour over a broad range of frequencies. It is certainly more beneficial to bring forth a theoretical framework that can accommodate such dependence. Kelvin-Voigt visco-elasticity has been used with some success, but this theory does not accommodate stress relaxation, which is critically important in polymers, when they undergo their glass transition, considered in either frequency or temperature space. Although for most materials this transition seems to occur in the lower frequency regime, one of the crucial aspects is that it affects both the real and imaginary part of the particular modulus \cite{liao2006estimation} whereas Kelvin-Voigt models only capture the latter.

In the present article, we return to the fundamentals of linear continuum mechanics and present a general, unified framework with which to model a variety of TVE materials of interest, with the specific interest in modelling how they  couple at interfaces. We discuss Kelvin-Voigt visco-elasticity and the standard linear models that extend this to incorporate stress relaxation. The same governing equations are used in any domain, without any need to identify the medium as a fluid, solid, viscoelastic, or otherwise. Needless to say parameters are required, but this means that a priori, all that is required is the identification of values that identify the medium as linear TVE, thus allowing one to model a vast range of important materials.

Over time the scientific community has developed a range of terms for specific media, e.g.\ \textit{visco-acoustic}, \textit{viscoelastic}, \textit{thermoelastic} etc.\ where certain physical effects are neglected. These are certainly useful and helpful because in many cases the neglected effects are not important. Here we also provide the asymptotic framework with which one can switch between these theories. In many cases it is straightforward and we simply set specific constants to zero, meaning that a lack of coupling arises. However in some cases one must be careful in the manner by which the theory is simplified, as we discuss. 

In Section \ref{sec:linearTVE} we begin with the conservation equations of homogeneous TVE materials, and describe local (in time) thermo-visco-elasticity before moving onto the more general non-local models that incorporate stress relaxation. These models are defined in the frequency domain and we consider Prony series that permit frequency dependence of material properties \cite{chen2000determining}. In Section \ref{sec:theorylimits} we go on to describe useful and appropriate asymptotic limits of the theory of thermo-visco-elasticity. Section \ref{sec:Application1:2HSs} covers the application of the various theories to the canonical problem of wave reflection from an interface between two continua, with the effects of coupling being illustrated and in particular the effects of relaxation on the frequency-dependent transmission and reflection. We close in Section \ref{sec:conclusions} with conclusions.

\begin{table}
\centering
\small
\begin{tabular}{ |p{6cm}||c|c|  }
 \hline
 \multicolumn{3}{|c|}{\textbf{Notation}} \\
 \hline
 Time derivative & \multicolumn{2}{|c|}{$\dot{\circ} = \partial \circ/\partial t$}  \\
 Gradient operator & \multicolumn{2}{|c|}{$\nabla \circ = \partial \circ/\partial x_i$} \\
 Laplacian operator &  \multicolumn{2}{|c|}{$\Delta \circ = \nabla \cdot \nabla \circ$} \\
 Tensor contraction &  \multicolumn{2}{|c|}{$\bm{A} \bm : \bm{B}= A_{ij}B_{ij}$} \\
 Matrix trace \& Matrix transpose & \multicolumn{2}{|c|}{$\tr(\circ)$ \& $\circ^T$} \\
  Fourier component & \multicolumn{2}{|c|}{${\circ} = \operatorname{Re}{\{\hat{\circ} \me^{-\mi \omega t}}\}$} \\
  Complex conjugate  & \multicolumn{2}{|c|}{$\circ^*$} \\
  Heaviside function & \multicolumn{2}{|c|}{$H$} \\
  Time average over wave period & \multicolumn{2}{|c|}{$\langle \circ \rangle = \frac{\omega}{2 \pi} \int_{0}^{{2\pi/\omega}} (\circ) \diff t$} \\
  Three-dimensional Identity tensor & \multicolumn{2}{|c|}{$\bm{I}$} \\
 \hline
 \multicolumn{3}{|c|}{\textbf{Thermo-Visco-Elastic {Parameters}}} \\
 \hline
 \textbf{Parameters} & \textbf{Unit(s)} & \textbf{Symbols and Definitions} \\
 \hline
 Continuum's displacement vector & m   & $\mathbf{u}$     \\
 Infinitesimal strain tensor & -- & $\bm{\varepsilon} = (\nabla \mathbf{u} + (\nabla \mathbf{u})^T )/2$\\
 Off-diagonal entries of the strain tensor & -- & $\bm{e}=\bm{\varepsilon} - \tr(\bm{\varepsilon}) \bm{I}/3$ \\
 Cauchy stress tensor & N m$^{-2}$ & $\bm{\sigma}$\\ 
 Off-diagonal entries of the stress tensor & N m$^{-2}$ & $\bm{s}=\bm{\sigma} - \tr(\bm{\sigma}) \bm{I}/3$ \\
 Linear and angular frequency &  Hz, rad s$^{-1}$ & $f$, $\omega$ \quad $\omega = 2\pi f$ \\
 Classical (isothermal) Lam\'e coefficients & N m$^{-2}$ & $\mu,\lambda >0 $\\
 Elastic bulk modulus (isothermal) & N m$^{-2}$   &$K = \lambda + 2 \mu/3$\\
 Elastic Poisson's ratio (isothermal) & - & $\nu$ \\
 Bulk and shear viscosity & N s m$^{-2}$ & $ \eta_K, \eta_\mu > 0, \quad \eta_\lambda = \eta_K -2 \eta_\mu/3 $ \\
 Viscosity parameter & N s m$^{-2}$ & $\zeta = 2 \eta_\mu + \eta_\lambda$\\
 Local in time complex Lam\'e quantities & N m$^{-2}$ & $\hat{ \lambda}= \lambda-\mi \omega \eta_\lambda$, \quad $\hat{ \mu} =\mu -\mi \omega \eta_\mu$ \\
 Local in time complex bulk modulus & N m$^{-2}$ & $\hat{K} = K -\mi \omega \eta_K$\\
 Thermal conductivity & W m$^{-1}$ K$^{-1}$ & $\mathscr{K}$\\
 Internal energy density per unit mass & N m kg$^{-1}$ & $\mathcal{E}$\\
 Total and ambient temperature & K  & $T, T_0$\\
 Non-dimensional temperature variation & -- & $\theta=(T-T_0)/T_0$ \\
 Total and ambient mass density &   kg m$^{-3}$  & $\rho, \rho_0$\\
 Total and ambient entropy per unit mass & N m kg$^{-1}$ K$^{-1}$ & $\mathfrak{s}, \mathfrak{s}_0$ \\
 Specific heat at constant pressure$/$volume& J kg$^{-1}$ K$^{-1}$  & $c_p, c_v \quad  \rho_0(c_p-c_v)=\alpha^2 K T_0$\\
 Ratio of specific heats & -- & $\gamma=c_p/c_v$ \\
 Adiabatic/isothermal acoustic speed of sound & m s$^{-1}$ & $c_\text{A}, c_\text{Iso}$ \quad  $c_\text{A} =\sqrt{\gamma} c_\text{Iso}$ \\
 Coefficient of thermal expansion & K$^{-1}$ & $\alpha$ \\
 Volumetric heat supply per unit mass & N m kg$^{-1}$ s$^{-1}$ & $B$ \\
 Body force per unit mass & N kg$^{-1}$ & $\mathbf{G}$\\
 Fourier-Stokes heat flux vector & N m$^{-1}$ s$^{-1}$ & $\mathbf{q}$\\
 Total and ambient Helmholtz free energy per unit mass & N m kg$^{-1}$ & $\Psi, \Psi_0$\\
 Thermal parameter &  m$^{-2}$ & $L_\phi$ \\
 Thermo-visco-elastic coupling quantity & -- & $L_\theta$\\
 Thermo-compressional wave-potentials \& wavenumbers  & m$^2$ \& m$^{-1}$  & $\varphi, \vartheta$  \quad \& \quad  $k_\varphi, k_\vartheta$ \\
 Pressure/Shear wave-potentials \& wavenumbers &  m$^2$ \&  m$^{-1}$ & $\phi, \mathbf{\Phi}$ \quad \& \quad $k_\phi, k_\Phi$\\
 Temperature contributions &  m$^{-2}$ & $\mathscr{T}_\varphi, \mathscr{T}_\vartheta$\\
 Mechanical relaxation functions &  N m$^{-2}$ & $\mathcal{R}_1, \mathcal{R}_2$\\
 Thermo-mechanical relaxation function & N m$^{-2}$ K$^{-1}$ & $ \mathcal{R}_3$\\
Specific heat relaxation function & J m$^{-3}$ K$^{-2}$ & $\mathcal{R}_4$\\
Energy flux vector per unit volume & N m$^{-1}$ s$^{-1}$ & $\mathbf{J}$\\
Total TVE energy per unit volume& N m$^{-2}$ & $\mathcal{U}$\\
Energy dissipation per unit time/volume & N m$^{-2}$ s$^{-1}$ & $\mathcal{D}$\\
 \hline
\end{tabular}
\caption{Thermo-viscous parameters and other quantities that appear in the general TVE model. By the "ambient" value of a quantity, we refer to its value prior to deformation, i.e. in the undeformed configuration which we assume to be still.}
\label{table: TVE parameters}
\end{table}

\section{Modelling linear TVE media}\label{sec:linearTVE}

\subsection{Governing equations}

Our starting point is the classical set of conservation laws of linear continuum mechanics: conservation of mass, momentum and energy, together with the Clausius-Duhem inequality \cite{marsden_mathematical_1994}
\begin{subequations}\label{Governing eqns 1}
\begin{align}
\label{eqn:MassBalance}
   \dot{\rho}+ \rho \nabla \cdot \dot{\mathbf{u}} &= 0, \\ \label{eqn:MomentumBalance}
\rho \ddot{\mathbf{u}} &= \nabla \cdot \bm{\sigma} + \rho \mathbf{G},\\ \label{eqn:EnergyBalance}
\rho \dot{\mathcal{E}}  + \nabla \cdot \mathbf{q} & = \bm{\sigma} \bm: \dot{\bm{\varepsilon}} + \rho B,\\ \label{eqn:2ndLawTherm}
 \rho T \dot{\mathfrak{s}} + \nabla \cdot \mathbf{q}  & \geq  \rho B + \frac{\mathbf{q} \cdot \nabla \mathbf{T}}{T},
\end{align}
\end{subequations}
where notation is summarised in Table \ref{table: TVE parameters}, and the symmetry of the Cauchy stress tensor $\bm{\sigma} = \bm{\sigma}^T$ arises due to conservation of angular momentum.

\subsection{Local (in time) TVE}\label{subsection:local TVE}

We assume that all constitutive models considered are local in space. We begin with the simplest (local) dependence on time, where we introduce the Helmholtz free energy per unit mass \cite{boley2012theory}
\begin{equation}\label{FreeEnergy Local theory}
    \Psi(t) \equiv \Psi(\bm{\varepsilon}(t), T(t))=\mathcal{E}(t) - T(t) \mathfrak{s}(t).
\end{equation}
Using this in (\ref{eqn:EnergyBalance}), (\ref{eqn:2ndLawTherm}) yields
\begin{equation}
    \left( \bm{\sigma} - \rho \frac{\partial  \Psi}{\partial \bm{\varepsilon}} \right):\dot{\bm{\varepsilon}}  - \left (\frac{\partial  \Psi}{\partial T} +\mathfrak{s}  \right) \rho \dot{T} - \frac{\mathbf{q} \cdot \nabla {T}}{T} \geq 0.
  \label{eqn:entropy-production}
\end{equation}
We then adopt the approach of \textit{Coleman-Noll} \cite{coleman1963thermodynamics} and \textit{Liu} \cite{liu1972method} to yield further information; since (\ref{eqn:entropy-production}) must hold for arbitrary deformations, the imposition of specific deformations permits conclusions to be deduced on functional form. A purely isothermal process ($\dot{T}=0, \nabla T = \mathbf{0}$) and a process that involves no deformation but a change in uniform temperature, respectively, yields
\begin{equation}\label{eqn: TE thermodynamic restrictions}
    \bm{\sigma}^{\text{TE}} = \rho \frac{\partial  \Psi}{\partial \bm{\varepsilon}}, \quad \text{and} \quad  
 \mathfrak{s} = - \frac{\partial  \Psi}{\partial T}, 
\end{equation}
where the superscript "TE" refers to \textit{thermo-elastic}. The conditions (\ref{eqn: TE thermodynamic restrictions}) are \textit{sufficient} but not \textit{necessary} to satisfy (\ref{eqn:entropy-production}): one can include an additional visco-elastic (VE) contribution to the Cauchy stress, e.g.\ for isotropic media
\begin{align}\label{C-Stress: TE+VE}
    \bm{\sigma}=  \bm{{\sigma}}^{\text{TE}}+ \bm{{\sigma}}^{\text{VE}} = {\rho} \frac{\partial  {\Psi}}{\partial \bm{\varepsilon}} + 2 {\eta}_\mu \dot{ \bm{\varepsilon}} + \left({\eta}_K - \frac{2}{3}{\eta}_\mu\right)  \tr(\dot{\bm{\varepsilon}}) \bm{I},
\end{align}
 where the shear and bulk viscosities\footnote{These terms are defined in several ways throughout the literature, our choice of ${\eta}_K$ as the bulk viscosity matches the convention of the elastic bulk modulus.} (both constants here) satisfy ${\eta}_\mu > 0, {\eta}_\lambda = {\eta}_K -2{\eta}_\mu/3 > 0$ and hence (\ref{C-Stress: TE+VE}) also satisfies (\ref{eqn:entropy-production}). The introduction of $\bm{\sigma}^{\text{VE}}$ distinguishes the current local-in-time TVE models from the commonly employed classical TE models. However, the absence of stress rate terms in (\ref{C-Stress: TE+VE}) is a strong restriction, since it fails to predict stress relaxation effects, which are important in many common materials such as polymers \cite{ferry1980viscoelastic}. Incorporating stress rates results in models that we refer to as \textit{non-local in time}, and this is the focus of Section \ref{sec:linearTVE}\ref{subsection: general TVE}. We first describe the thermal constitutive models and then the associated equations that describe local-in-time TVE wave propagation.
 
 We adopt Fourier's law of heat conduction,
\begin{equation}\label{Fouriers heat law local TVE}
    \mathbf{q} = -\mathscr{K} \nabla T,
\end{equation}
where ${\mathscr{K}} >0$ is the thermal conductivity of the material, whose positivity ensures that the last term in (\ref{eqn:entropy-production}) is never negative. The form (\ref{Fouriers heat law local TVE}) is the simplest admissible choice, resulting in a parabolic diffusion equation (\ref{eqn:Linearheatflow}) for which the thermal wave-speed is infinite. Thermal waves with finite velocity (e.g.\ Maxwell-Cattaneo heat waves) are obtained when introducing a thermal relaxation time which arises when taking into account the rate of heat flux vector in (\ref{Fouriers heat law local TVE}) \cite{cattaneo1958form,lord1967generalized}.

At this stage it only remains to determine the thermodynamically consistent form of ${\Psi}$. As shown in Appendix \ref{local TVE derivation}, for a (local) linear theory of TVE we obtain
\begin{subequations}\label{eqn:stress and entropy}
\begin{align}\label{stressTensorFinal}
     \bm{\sigma} &= 2{\mu} \bm{\varepsilon} + 2{\eta}_\mu\bm{\dot{\varepsilon}} +({\lambda} \tr (\bm{\varepsilon})+{\eta}_\lambda \tr (\dot{\bm{\varepsilon}})-{\alpha} {K} {T}_0 \theta)\bm{I},\\
      {\mathfrak{s}} &= {\mathfrak{s}}_0 + {c}_v \theta +\frac{{\alpha} {K}}{{\rho}_0} \tr (\bm{\varepsilon}),\label{entropyFinal}
\end{align}
\end{subequations}
where $\theta=({T}-{T}_0)/{T}_0$ is the non-dimensional temperature difference, and
${K} = {\lambda} + 2{\mu}/3$ denotes the elastic bulk modulus measured at a state of constant temperature (i.e. isothermal like $\lambda$ and $\mu$, see Appendix \ref{local TVE derivation}). Note that by introducing the off-diagonal tensors $\bm{s}$ and  $\bm{e}$ satisfying
\begin{equation}
\label{linearStresstensor decomposed}
    \bm{s} = \bm{\sigma} -  \frac{1}{3} \tr(\bm{\sigma}) \bm{I}, \qquad
    \bm{e} = \bm{\varepsilon} - \frac{1}{3} \tr(\bm{\varepsilon}) \bm{I},
\end{equation}
we may deduce from (\ref{stressTensorFinal}) that
\begin{equation}\label{localTVE: sij and trace}
    \bm{s} = 2{\mu} \bm{e} + 2{\eta}_\mu\bm{\dot{e}}, \qquad \tr{(\bm{\sigma})}=3 \left[K \tr{(\bm{\varepsilon})}  + \eta_K \tr{(\bm{\dot{\varepsilon}})} - \alpha K T_0 \theta \right].
\end{equation}
Substituting (\ref{FreeEnergy Local theory}), (\ref{C-Stress: TE+VE}) and (\ref{entropyFinal}) into (\ref{eqn:EnergyBalance}) yields the energy equation
\begin{equation}\label{eqn:Linearheatflow}
    \mathscr{K} \Delta \theta - {\rho}_0 {c}_v \dot{\theta} = \alpha K \tr (\dot{\bm{\varepsilon}}),
\end{equation}
where we have assumed no external heat supply such that ${B}=0$. Note that viscous effects are not explicit in (\ref{eqn:Linearheatflow}) since they are quadratic in $\dot{\bm{\varepsilon}}$, and hence at this order the energy is analogous to that of linear thermo-elasticity \cite{boley2012theory}. Finally for convenience we write 
\begin{equation}\label{eqn:thermodIdentity}
     {\rho}_0(\gamma-1)=\frac{{\alpha}^2{K} {T}_0}{{c}_v},
\end{equation}
where $\gamma = {c}_p/{c}_v$ denotes the ratio of specific heats. Equation (\ref{eqn:thermodIdentity}) is a classical conserved quantity in thermo-elasticity (see Appendix \ref{local TVE derivation}). It is useful in practice since for solids ${c}_v$ is difficult to measure as opposed to ${c}_p$. As we see shortly  it plays an important role when considering the thermo-visco-acoustic (TVA) limit.
Finally, we show in electronic supplementary material Section SM1 that for this model we can obtain the \textit{energy conservation-dissipation corollary}
\begin{equation}\label{energy-conserv-diss}
    {\nabla} \cdot {\mathbf{J}} +  \frac{1}{2}{\dot{\mathcal{U}}} = - {\mathcal{D}}, \quad \begin{cases}
    &{\mathbf{J}} = - (\boldsymbol{{\sigma}}^\text{TVE}  \dot{{\mathbf{u}}} + {T}_0 \theta  {\mathscr{K}}{\nabla} \theta),\\ &{\mathcal{U}} = 
         {\rho}_0 |\dot{{\mathbf{u}}}|^2 + {\rho}_0 {T}_0 {c}_v \theta^2 + 2{\mu} |\bm \varepsilon|^2  + {\lambda}(\tr \bm \varepsilon )^2, \\
         &{\mathcal{D}}  = {T}_0 {\mathscr{K}} |{\nabla} \theta|^2 + 2 \eta_\mu |\dot{\bm \varepsilon}|^2
    + {\eta}_\lambda \left( \tr \dot{\bm \varepsilon} \right)^2,
    \end{cases}
\end{equation}
where $\mathbf{J}$ represents the energy flux vector, $\mathcal{D}$ the energy dissipation and $\mathcal{U}$ the total TVE energy. Note that ${\mathcal{D}} \geq 0$, so that a non-zero temperature gradient and strain rate always dissipates energy. Similar results are given for visco-elasticity in \cite[p. 20]{borcherdt2009viscoelastic} and for thermo-elasticity in \cite{caviglia2005harmonic}.

\subsubsection{Frequency domain decomposition for the local-in-time equations}\label{subsec:TVEwaveEQNS}

We now assume time-harmonic propagation of the form $\{{\mathbf{u}}, \theta, \bm{\sigma}\}({\mathbf{x}},{t})=\operatorname{Re}{\{\{ {\hat{\mathbf{u}}},\hat{\theta},\hat{\bm{\sigma}}\}({\mathbf{x}})\me^{-\mi {\omega} {t}} }\}$ and define the complex-valued (Kelvin-Voigt like) Lam\'e parameters 
\begin{equation}\label{eqn:Kelvin-voigt complex NL parameters}
   { \hat{\lambda}} = {\lambda} - \mi {\omega} {\eta}_{\lambda} \quad \text{and} \quad {\hat{\mu}} = {\mu} - \mi {\omega} {\eta}_{\mu}, 
\end{equation}
so that with (\ref{eqn:Kelvin-voigt complex NL parameters}) we can write the time-harmonic Cauchy stress (\ref{stressTensorFinal}) as
\begin{equation}\label{eqn:stressTEnsor}
 {\hat{\bm{\sigma}}} = \left({\hat{\lambda}} \nabla \cdot \hat{\mathbf{u}} -{\alpha} {K} {T}_0\hat{\theta}\right)\bm{I}+{\hat{\mu}}\left(\nabla \hat{\mathbf{u}} + (\nabla \hat{\mathbf{u}})^T \right),
\end{equation}
or equivalently with $\{ \bm{s},\bm{e}\}({\mathbf{x}},{t})=\operatorname{Re}{\{\{ \hat{\bm{s}},\hat{\bm{e}}\}({\mathbf{x}})\me^{-\mi {\omega} {t}} }\}$ (\ref{localTVE: sij and trace}) becomes
\begin{equation}\label{local TVE: sij and trace freq dom}
  {\hat{\bm{s}}} = 2{\hat{\mu}} \hat{\bm{e}}, \qquad
    {\tr(\hat{\bm{\sigma}})} = 3\left[(K - \mi \omega \eta_K) \tr(\hat{\bm{\varepsilon}}) - \alpha K {T}_0 \hat{\theta}\right].
\end{equation}
Substituting (\ref{eqn:stressTEnsor}) in the conservation of momentum equation (\ref{eqn:MomentumBalance}) yields
\begin{align}\label{eqn:NavierLameForced}
     ({\hat{\lambda}} + 2{\hat{\mu}}) \mathbf{{\nabla}} \left( \mathbf{{\nabla}} \cdot {\hat{\mathbf{u}}}\right) - {\hat{\mu}} \mathbf{{\nabla}} \times \mathbf{{\nabla}} \times {\hat{\mathbf{u}}} + {\rho}_0 {\omega}^2  {\hat{\mathbf{u}}} = {\alpha} {K} {T}_0 \mathbf{\nabla}  \hat{\theta},
\end{align}
which corresponds to Navier-Lam{\'e} with thermo-mechanical coupling as in classical linear TE. Introducing the classical Helmholtz potentials $\phi, \mathbf{\Phi}$ in the form
\begin{align}\label{eqn:diplacementDecompos}
    {\hat{\mathbf{u}}} &= {\mathbf{\nabla}} {\phi} + {\mathbf{\nabla}} \times  {\mathbf{\Phi}}, & {\nabla} \cdot {\mathbf{\Phi}} &= 0,
\end{align}
and making use of Helmholtz' theorem \cite{graff2012wave}, we deduce that the potentials must satisfy
\begin{subequations}\label{Local TVE coupled PDEs}
\begin{align}\label{eqn:coupledphi}
    \Delta {\phi} +{k}_\phi^2{\phi}+L_\theta\hat{\theta}&=0, \\ \label{eqn:decoupledCURLpart}
     \Delta {\mathbf{\Phi}} + {k}_\Phi^2 {\mathbf{\Phi}} &=\mathbf{0}, \\\label{eqn:coupledTau}
     \Delta \hat{\theta} +{k}_{\theta}^2\hat{\theta}+{L}_\phi \Delta {\phi}&= 0,
\end{align}
\end{subequations}
where
\begin{align}\label{eqn:TVEwavenumbers}
    {k}_{\theta}^2=\mi {c}_v\frac{{\rho}_0 {\omega}}{{\mathscr{K}}}, \quad {k}_{\phi}^2=\frac{{\rho}_0 {\omega}^2}{{\hat{\lambda}}+2{\hat{\mu}}}, \quad {k}_{\Phi}^2=\frac{{\rho}_0 {\omega}^2}{{\hat{\mu}}},
\end{align}
and where we have defined ${L}_\phi$ (with dimension $\si{\meter}^{-2}$), and the non-dimensional thermo-mechanical coupling parameter $L_\theta$ as
\begin{equation}\label{eqn:lphi and ltau}
    {L}_\phi=\frac{\mi {\alpha} {K} {\omega}}{{\mathscr{K}}}, \quad \text{and} \quad L_\theta=-\frac{{\alpha}  {T}_0 {K}}{{\hat{\lambda}}+2{\hat{\mu}}}.
\end{equation}
In the limit ${\alpha} \rightarrow 0$ the system (\ref{Local TVE coupled PDEs}) uncouples immediately. Moreover, it is the size of $|L_\theta|$ that determines the importance of thermo-elastic coupling. In order to obtain a less restrictive theory, it is often argued for many materials in common scenarios that $|{L}_\phi|\times (\text{"characteristic length"})^2 \ll 1$ so that the energy equation (\ref{eqn:coupledTau}) becomes uncoupled. The corresponding solution can then be fed into (\ref{eqn:coupledphi}) to obtain a forced Helmholtz equation with a known source term. This approximation is sometimes referred to as the \textit{theory of thermal stresses} in order to distinguish it from fully coupled thermo-elasticity  \cite{nowacki2013thermoelasticity}.

\noindent To decouple the system (\ref{Local TVE coupled PDEs}) completely substitute (\ref{eqn:coupledphi}) into (\ref{eqn:coupledTau}) to obtain
\begin{align}\label{eqn:order4}
    {\mathcal{L}}_O \lbrace{{\phi}\rbrace} &=0, \qquad \text{where} \qquad {\mathcal{L}}_O = (\Delta+({a}-{b}))(\Delta+({a}+{b})),\\ \label{eqns: dimensional a b wavenumbers }
    {a}&=\frac{1}{2}\left({k}_\theta^2+{k}_\phi^2-L_{\theta}{L}_{\phi} \right), \quad \text{and} \quad {b} =\sqrt{{a}^2-{k}_\phi^2{k}_\theta^2}.
\end{align}
The solution to (\ref{eqn:order4}) is thus equivalent to solving the pair of Helmholtz equations
\begin{subequations}\label{TVE LOcal 2 bulk PDEs}
\begin{align}\label{eqn:varthetahelmholtz}
 \Delta {\vartheta} +{k}_\vartheta^2 {\vartheta}&=0,\\ \label{eqn:varphihelmholtz}
    \Delta {\varphi}+{k}_\varphi^2 {\varphi}&=0,
\end{align}
\end{subequations}
with
\begin{equation} \label{eqn:DimTVEdecoupled wavenumbers kvarphi kvartheta}
   {k}_\vartheta^2 = {a} + {b}, \quad {k}_\varphi^2 = {a} - {b} .
\end{equation}
Employing (\ref{eqn:coupledphi}), the two newly introduced potentials $\varphi, \vartheta$ are related to $\phi$ and $\hat{\theta}$ via the matrix form\footnote{Due to the uniqueness of the solution to the linear PDE (\ref{eqn:order4}) being up to a constant, we may also write $\phi = C_1 \varphi + C_2\vartheta$, whence $ \hat{\theta} L_\theta = C_1({k}_{\varphi}^2-{k}_{\phi}^2){\varphi} + C_2({k}_{\vartheta}^2-{k}_{\phi}^2){\vartheta}$, for constants $C_1, C_2$ but here we choose $C_1 = C_2 = 1$ to match the conventional approach.}

\begin{equation} \label{eqn:varphithetamatform}
    \begin{pmatrix}
	{\phi} \\
	\hat{\theta}
	\end{pmatrix} =
		\left(
		\begin{array}{cc}
		 1         &       1 \\
		 {\mathscr{T}}_\varphi & {\mathscr{T}}_\vartheta  \\
		\end{array}
		\right)
	\begin{pmatrix}
	{\varphi} \\
	{\vartheta}
	\end{pmatrix},
\end{equation}
where
\begin{equation}\label{temperature contributions}
    {\mathscr{T}}_\varphi=\frac{1}{L_{\theta}}(k_\varphi^2 - {k}_\phi^2)  , \quad {\mathscr{T}}_\vartheta =\frac{1}{L_{\theta}}(k_\vartheta^2 - {k}_\phi^2).
\end{equation}
As is well known therefore, the equations of motion for linear local-in-time TVE are thus governed by the three Helmholtz equations (\ref{eqn:decoupledCURLpart}), (\ref{eqn:varthetahelmholtz}) and (\ref{eqn:varphihelmholtz}) from which we can recover the temperature and displacement fields through (\ref{eqn:diplacementDecompos}), and (\ref{eqn:varphithetamatform}). These wave potentials consist of two thermo-compressional potentials ${\varphi}, {\vartheta}$ and a shear potential ${\mathbf{\Phi}}$, the latter being indepedent of thermal effects. They can be directly correlated to those of \cite{deschamps1989liquid} as well as \cite{iecsan2011theory} (by taking the limit of zero volume fraction of voids). Asymptotic approximations to (\ref{eqn:DimTVEdecoupled wavenumbers kvarphi kvartheta}), (\ref{temperature contributions}) and their validity will be discussed in Section \ref{sec:linearTVE}\ref{subsection: asymptotics nonlocalTVE} but before we move on to incorporating the influence of stress relaxation.

Finally, as we will see later, it is useful to represent the intensity of time-harmonic waves as an average of $\mathbf{J}$ (\ref{energy-conserv-diss}) over the wave period ($2\pi/\omega$) such that
\begin{equation} \label{eqn: time average energy flux}
     \langle \mathbf{{J}}\rangle (\mathbf{x},\omega) = \frac{\omega}{{2\pi}} \int_{0}^{{2\pi/\omega}} \mathbf{{J}} (\mathbf{x},t) \diff t = -\frac{1}{2} \operatorname{Re} \{\bm{{\sigma}} \dot{{\mathbf{u}}}^{*}  + \theta {\mathscr{K}} {\nabla} \theta^* \},
\end{equation}
where asterisk $^*$ denotes complex conjugate. Equation (\ref{eqn: time average energy flux}) represents the average energy flux (per unit area) due to both the mechanical power and the heat flux, see e.g. \cite{caviglia2005harmonic}.
\subsection{Non-local (in time) TVE: the influence of stress relaxation}\label{subsection: general TVE}

The local-in-time TVE constitutive model (\ref{stressTensorFinal}) has no dependence on \textit{history}, or equivalently as it turns out, no information with regard to stress rates. Whilst the model as presented permits the modelling of \textit{creep compliance}, it means that \textit{stress relaxation} cannot be modelled. From a physical viewpoint, this limits its applicability, especially for the diverse range of polymeric materials in which relaxation, or equivalently, strongly frequency-dependent material properties, is common. In order to accommodate this effect \textit{and} creep, we must take into consideration the kinematical and thermal \textit{time histories}, so that the Helmholtz free energy per unit mass now takes the form
\begin{equation}\label{TVE:Free energy General}
    {\Psi} \equiv {\Psi} ( \left. \bm{\varepsilon}(\tau)\right|_{\tau = - \infty}^t,\left. T(\tau)\right|_{\tau = - \infty}^t).
\end{equation}
This makes the question of whether the Clausius-Duhem inequality (\ref{eqn:2ndLawTherm}) is solved less trivial, even for linear theories \cite{christensen1967linear,christensen2012theory}. Instead, following \cite{christensen1967linear} the equations for a linear isotropic medium (\ref{eqn:stress and entropy}), (\ref{localTVE: sij and trace}) generalise to
\begin{subequations}\label{TVE Non Local Eqns}
\begin{align}\label{eqn:DeviatoricStressChristensen}
 \bm{s} &= \int_{- \infty}^{{t}}  {\mathcal{R}}_1({t}-{\mathcal{T}}) \dot{\bm{e}}(\mathcal{T}) \diff {\mathcal{T}}, \\ \label{eqn:DiagonalStressChristensen}
    \tr{(\bm{\sigma})} &= \int_{- \infty}^{{t}} {\mathcal{R}}_2({t}-{\mathcal{T}}) \tr(\dot{\bm{\varepsilon}}({\mathcal{T}})) \diff {\mathcal{T}} - 3 {T}_0\int_{- \infty}^{{t}} {\mathcal{R}}_3({t}-{\mathcal{T}}) \dot{\theta}({\mathcal{T}}) \diff {\mathcal{T}} , \\ \label{entropy a la Christensen time domain} 
   {\rho}_0 \mathfrak{s} &=  {\rho}_0  {\mathfrak{s}}_0 +   {T}_0 \int_{- \infty}^{{t}}  {\mathcal{R}}_4({t}-{\mathcal{T}}) \dot{\theta}({\mathcal{T}}) \diff {\mathcal{T}} + \int_{- \infty}^{{t}} {\mathcal{R}}_3({t}-{\mathcal{T}}) \tr(\dot{\bm{\varepsilon}}({\mathcal{T}})) \diff {\mathcal{T}},
\end{align}
\end{subequations}
and the energy equation (\ref{eqn:Linearheatflow}) is replaced by
\begin{equation}
    \label{eqn:IsotropicEnergyChristensen}
{\mathscr{K}} \Delta \mathcal{\theta} = \frac{\partial}{\partial {t}}\left(T_0 \int_{- \infty}^{{t}} {\mathcal{R}}_4({t}-{\mathcal{T}}) \dot{\theta}({\mathcal{T}}) \diff {\mathcal{T}} + \int_{- \infty}^{{t}} {\mathcal{R}}_3({t}-{\mathcal{T}}) \tr(\dot{\bm{\varepsilon}}({\mathcal{T}})) \diff {\mathcal{T}}\right),
\end{equation}
where the kernels ${\mathcal{R}}_1,{\mathcal{R}}_2,{\mathcal{R}}_3, {\mathcal{R}}_4$ are relaxation functions\footnote{Here these functions are scalar valued since we are only considering isotropic deformations.} containing the time varying thermo-mechanical properties of the medium. Note that despite including thermal history in the present theory, general thermodynamic consistency again requires Fourier's law (\ref{Fouriers heat law local TVE}) to hold with a constant thermal conductivity $\mathscr{K}$ \cite{christensen1967linear}, so that (\ref{eqn:IsotropicEnergyChristensen}) remains parabolic. 

Restrictions on ${\mathcal{R}}_i$ include causality, giving ${\mathcal{R}}_i({\tau}) = 0$ for ${\tau} < 0$ (where $i=$1$-$4) and choosing a form such that all integrals in (\ref{TVE Non Local Eqns}), (\ref{eqn:IsotropicEnergyChristensen}) are convergent. Finally, the choice must satisfy the \textit{dissipation inequality}\footnote{Unlike in the local case above where this inequality is automatically satisfied by setting constant valued viscosities, the generality of time non-locality implies an extra restriction on the relaxation functions.}:
\begin{equation}
    {\Lambda} \geq 0, \label{intro:Lambda geq 0}
\end{equation}
where ${\Lambda}$ is detailed in electronic supplementary material Section SM2. Despite being frequently neglected, the requirement (\ref{intro:Lambda geq 0}) is also present in the analogue isothermal VE theory. For a particular choice of ${\mathcal{R}}_i$, it can in principle be checked whether (\ref{intro:Lambda geq 0}) is satisfied. We note that the thermodynamics of certain widely used theories are often unclear, as in the case of Fung's Quasilinear visco-elasticity (QLV) theory  \cite{berjamin2021thermodynamic}. Unless otherwise stated, in the subsequent work we assume that we meet the necessary requirements for equations (\ref{TVE Non Local Eqns})-(\ref{eqn:IsotropicEnergyChristensen}) to apply.\\

In general it is non-trivial to determine the time-dependent form of the relaxation functions for a given material. They are assumed to depend only on the background temperature (assumed constant) ${T}_0$ as any more general temperature dependence must involve non-linearities, which are outside the scope of this paper. An exception is given by "thermo-rheologically simple" materials \cite{hunter1961tentative}, where the dependence of the material properties on temperature has a particularly appealing structure that allows for description with a linear theory. The dependence of these properties on temperature can be associated with a shift of the behaviour at a base constant temperature which is commonly known as the \textit{"time-temperature superposition"}. The particular shift function can in general be found experimentally but a very common empirical shift function is that of the Williams–Landel–Ferry \cite{ferry1980viscoelastic}.

Having established a sufficiently general constitutive framework with which to model materials with time-dependent material properties we now discuss how this can be described in the frequency domain. 

\subsubsection{Frequency domain decomposition for the non-local equations}

Assume now that the fields are time-harmonic, of the form
\begin{equation}\label{eqn:DevStressTimeHarmonic}
    \{\mathbf{u},\theta,\bm{\sigma},\bm{s},\bm{\varepsilon},\bm{e} \}({\mathbf{x},t)}= \operatorname{Re}{\{ \{\hat{\mathbf{u}},\hat{\theta},\hat{\bm{\sigma}},\hat{\bm{s}},\hat{\bm{\varepsilon}},\hat{\bm{e}} \}({\mathbf{x})}\me^{-\mi {\omega} {t}}\}},
\end{equation}
and decompose all relaxation functions as
\begin{equation}\label{relaxationFunction decomposed}
    {\mathcal{R}_i}(t) = {\mathcal{R}_i}^\prime + {\mathscr{R}_i}(t), \qquad \text{s.t.} \qquad {\mathscr{R}}_i(t) \rightarrow 0 \quad \text{as} \quad t\rightarrow \infty,
\end{equation}
for $i=1,2,3,4$ where ${\mathcal{R}_i}^\prime$ denotes the long-time asymptote, in the limit $t \to \infty$, and ${\mathscr{R}}_i(t)$ is the time dependent part.  This makes the treatment of the frequency transforms below simpler \cite{christensen2012theory}.

We can then substitute  (\ref{eqn:DevStressTimeHarmonic}) with (\ref{relaxationFunction decomposed}) into (\ref{TVE Non Local Eqns}) to obtain
\begin{equation}\label{nonlocalTVE: sij and trace freq dom}
  {\hat{\bm{s}}} = 2{\tilde{\mu}}(\mi \omega) \hat{\bm{e}}, \qquad
    {\tr(\hat{\bm{\sigma}})} = 3\left[{\tilde{K}}( \mi {\omega}) \tr(\hat{\bm{\varepsilon}}) - {T}_0 {\tilde{\mathcal{R}}}_3(\mi {\omega})\hat{\theta}\right],
\end{equation}
which may be directly compared to (\ref{local TVE: sij and trace freq dom}). In order to write (\ref{nonlocalTVE: sij and trace freq dom}) we have defined
\begin{subequations}\label{3complexRelaxationFuns}
\begin{align} \label{eqn:ComplexshearRelaxation}
   & {\tilde{\mu}}(\mi \omega) =\frac{1}{2}\left({\mathcal{R}}_1^\prime - \mi {\omega} \int_{0}^{\infty} {\mathscr{R}}_1({\mathcal{V}})\me^{\mi {\omega} {\mathcal{V}}} \diff {\mathcal{V}} \right), \quad 
  {\tilde{K}}(\mi \omega) = \frac{1}{3}\left({\mathcal{R}}_2^\prime - \mi {\omega} \int_{0}^{\infty} {\mathscr{R}}_2({\mathcal{V}})\me^{\mi {\omega} {\mathcal{V}}} \diff {\mathcal{V}}\right),\\ \label{eqn:ComplexzetaRelaxation}
& \phantom{asdadasdasdasdasaadd} {\tilde{\mathcal{R}}}_3(\mi {\omega}) = {\mathcal{R}}_3^\prime -\mi {\omega} \int_{0}^{\infty} {\mathscr{R}}_3({\mathcal{V}})\me^{\mi {\omega} {\mathcal{V}}} \diff {\mathcal{V}} ,
 \end{align}
\end{subequations}
which respectively corresponds to the complex shear modulus, the three-dimensional complex bulk modulus, and the complex modulus associated with the coefficient of thermo-mechanical coupling. From (\ref{eqn:ComplexshearRelaxation}) it follows that we can define the generalized first Lam\'e modulus, Poisson's ratio and Young's modulus, respectively as \cite{tschoegl2002poisson}
\begin{equation}\label{eqn:nonlocalTVE lambda}
    {\tilde{\lambda}}( \mi {\omega}) = {\tilde{K}}( \mi {\omega}) - \frac{2}{3}{\tilde{\mu}}( \mi {\omega}),
\qquad  {\tilde{\nu}}( \mi {\omega}) = \frac{3{\tilde{K}}( \mi {\omega}) - 2 {\tilde{\mu}}( \mi {\omega})}{6 {\tilde{K}}( \mi {\omega}) + 2 {\tilde{\mu}}( \mi {\omega})},\qquad  {\tilde{E}}( \mi {\omega}) = \frac{9 \tilde{K}( \mi {\omega}) {\tilde{\mu}}( \mi {\omega}) }{3{\tilde{K}}( \mi {\omega}) +  {\tilde{\mu}}( \mi {\omega})}. 
\end{equation}
Finally, using (\ref{eqn:ComplexzetaRelaxation}) the energy balance equation (\ref{eqn:IsotropicEnergyChristensen}) becomes
\begin{equation}
    {\mathscr{K}} \Delta \hat{\theta} + \mi {\omega}  \left({T}_0 {\tilde{\mathcal{R}}}_4(\mi \omega) \hat{\theta} + {\tilde{\mathcal{R}}}_3(\mi {\omega})\tr(\hat{\bm{\varepsilon}}) \right) = 0,
\end{equation}
where we defined the complex modulus
\begin{equation}\label{eqn:ComplexmRelaxation}
    {\tilde{\mathcal{R}}}_4(\mi {\omega}) = {\mathcal{R}}_4^\prime - \mi {\omega} \int_{0}^{\infty} {\mathscr{R}}_4({\mathcal{V}})\me^{\mi {\omega} {\mathcal{V}}} \diff {\mathcal{V}} .
\end{equation}
In fact, by direct comparison with the energy equation commonly used in linear thermo-elasticity (e.g.  (1.12.22) in \cite{boley2012theory}) we observe that this quantity can be interpreted as a  specific heat at constant strain/volume per unit volume, which in the setting of TVE with temperature history allows for frequency dependency, i.e.
\begin{equation}\label{eqn:specificheatrelaxation}
   {\tilde{\mathcal{R}}}_4(\mi {\omega}) =\frac{{\rho}_0}{{T}_0} {\tilde{c}}_v(\mi {\omega}).
\end{equation}
Hence, the associated Cauchy stress in the frequency domain becomes 
\begin{align}
    \label{eqn:timeharmonicstressTVE}
    {\hat{\bm{\sigma}}}^{\text{TVE}} &= 2{\tilde{\mu}}(\mi {\omega})\hat{\bm{e}} + ({\tilde{K}}(\mi \omega)\tr(\hat{\bm{\varepsilon}}) - {T}_0 \tilde{\mathcal{R}}_3(\mi {\omega})\hat{\theta})\bm{I} = 2{\tilde{\mu}}(\mi {\omega}) \hat{\bm{\varepsilon}} + ({\tilde{\lambda}}( \mi {\omega})\tr(\hat{\bm{\varepsilon}})  - {T}_0 {\tilde{\mathcal{R}}}_3(\mi {\omega})\hat{\theta})\bm{I}.
\end{align}
The associated energy and momentum equations reduce to
\begin{subequations}\label{nonlocalTVE stress and energy with relaxation}
\begin{align}
   \label{eqn:timeharmonicenergyTVE}
    &{\mathscr{K}} \Delta \hat{\theta} + \mi {\omega} {\rho}_0  {\tilde{c}}_v(\mi {\omega}) \hat{\theta} +  \mi {\omega} {\tilde{\mathcal{R}}}_3(\mi {\omega}){\mathbf{\nabla}} \cdot {\hat{\mathbf{u}}} = 0,\\ \label{eqn:timeharmonicmomentumTVE}
   & ({\tilde{\lambda}}( \mi {\omega}) + 2{\tilde{\mu}}(\mi {\omega})) {\mathbf{\nabla}} \left({\mathbf{\nabla}} \cdot {\hat{\mathbf{u}}} \right) - {\tilde{\mu}}(\mi {\omega}) {\mathbf{\nabla}} \times {\mathbf{\nabla}} \times  {\hat{\mathbf{u}}} - {T}_0{\tilde{\mathcal{R}}}_3(\mi {\omega}) {\mathcal{\nabla}} \hat{\theta} +{\rho}_0 {\omega}^2 {\hat{\mathbf{u}}} = \mathbf{0}.
\end{align}
\end{subequations}
Conveniently, equations (\ref{eqn:timeharmonicstressTVE})-(\ref{eqn:timeharmonicmomentumTVE}) have the same structure as (\ref{eqn:Kelvin-voigt complex NL parameters})-(\ref{eqn:NavierLameForced}), (\ref{Local TVE coupled PDEs}), for a fixed frequency $\omega$. Now however, rich frequency dependent behaviour can be accommodated by the incorporation of the relaxation functions. The equivalent form however means that the decomposition of Section \ref{sec:linearTVE}\ref{subsection:local TVE}\ref{subsec:TVEwaveEQNS} remains valid so that the fields remain solutions of the decoupled Helmholtz equations (\ref{eqn:decoupledCURLpart}), (\ref{eqn:varthetahelmholtz}) and (\ref{eqn:varphihelmholtz}), as in the local case, with the only change (but a critical one) being that now the relevant quantities appearing in the wavenumbers have a more general frequency dependence:
\begin{subequations}\label{eqns:kdiff}
\begin{align} 
    {k}_{\theta}^2= \mi {\tilde{c}}_v(\mi {\omega})\frac{{\rho}_0 {\omega}}{{\mathscr{K}}}, \quad {k}_{\phi}^2=\frac{{\rho}_0 {\omega}^2}{{\tilde{\lambda}}(\mi {\omega})+2{\tilde{\mu}}(\mi {\omega})}, \quad {k}_{\Phi}^2=\frac{{\rho}_0 {\omega}^2}{{\tilde{\mu}}(\mi {\omega})}, \label{DIMTVE: general elements of wavenumber with memory1}\\
    \quad {L}_\phi = \frac{\mi {\omega} {\tilde{\mathcal{R}}}_3(\mi {\omega})}{{\mathscr{K}}}, \quad L_\theta = - \frac{{T}_0 {\tilde{\mathcal{R}}}_3{(\mi {\omega})}}{{\tilde{\lambda}}(\mi {\omega})+2{\tilde{\mu}}(\mi {\omega})}. \label{DIMTVE: general elements of wavenumber with memory2}
\end{align}
\end{subequations}
As in the local case, the displacement and temperatures are given respectively by appropriate combinations of the potentials:
\begin{subequations}\label{eqns:TVE w relaxation}
\begin{align}\label{TVE w relaxation displacement}
    {\hat{\mathbf{u}}}^{\text{TVE}} &=  {\nabla}( {\vartheta} + {\varphi}) +  {\nabla}  \times {\mathbf{\Phi}}, \\ \label{TVE w relaxation temperature}
    \hat{\theta}^{\text{TVE}} &= \frac{1}{L_{\theta}}({a} - {b} - {k}_\phi^2) {\varphi} + \frac{1}{L_{\theta}}({a} + {b} - {k}_\phi^2) {\vartheta}.
\end{align}
\end{subequations}
Note, in particular, that the more general frequency dependence of the TVE coupling parameter $L_\theta$ shows that certain materials may exhibit significant thermal coupling only for certain frequency ranges.

\subsubsection{Form of relaxation functions}\label{subsection:Choice of moduli}

\begin{figure}
\centering
\begin{subfigure}{.5\textwidth}
  \centering
  \includegraphics[width=1\linewidth]{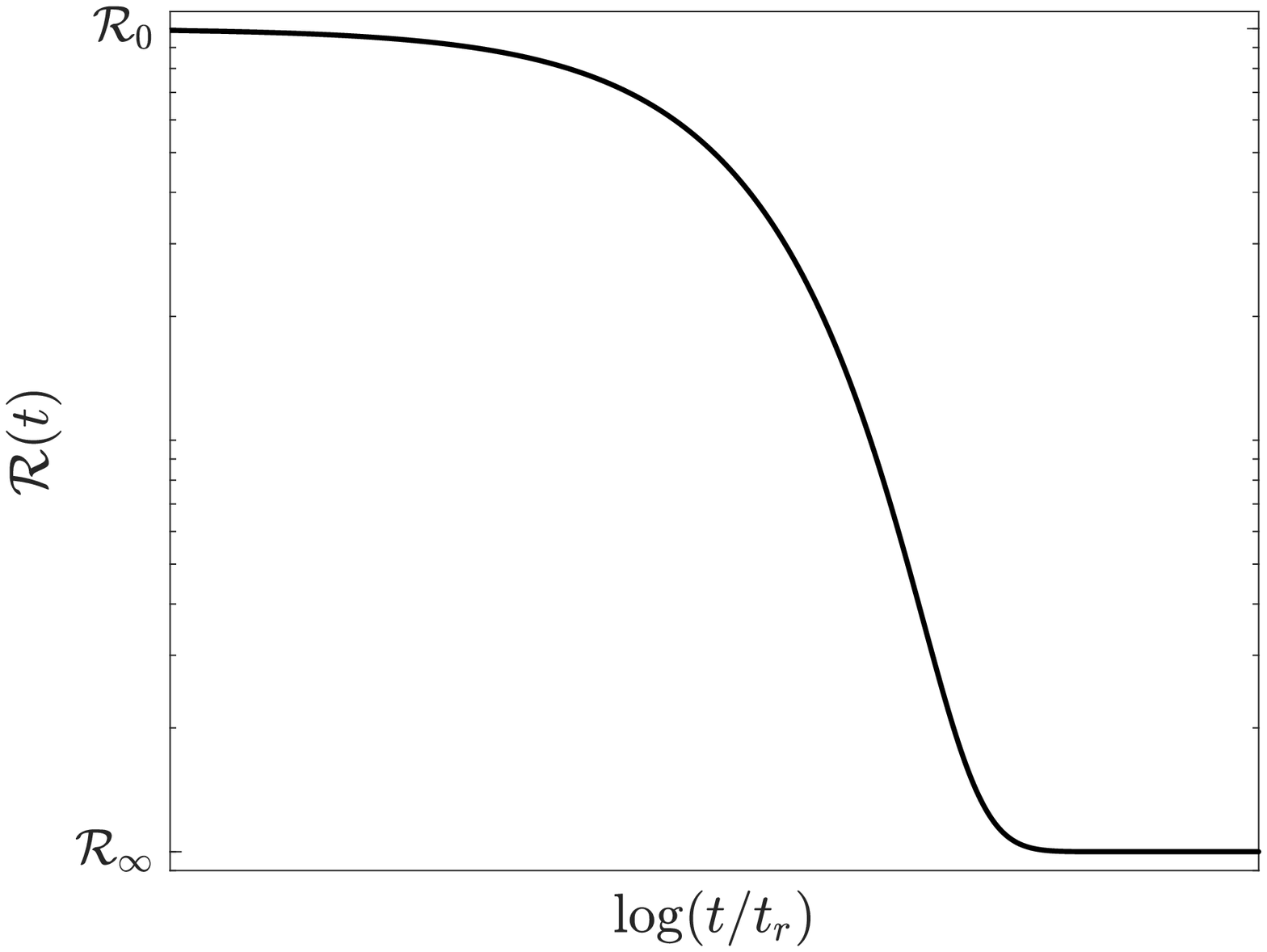}
\end{subfigure}%
\begin{subfigure}{.5\textwidth}
  \centering
  \includegraphics[width=1\linewidth]{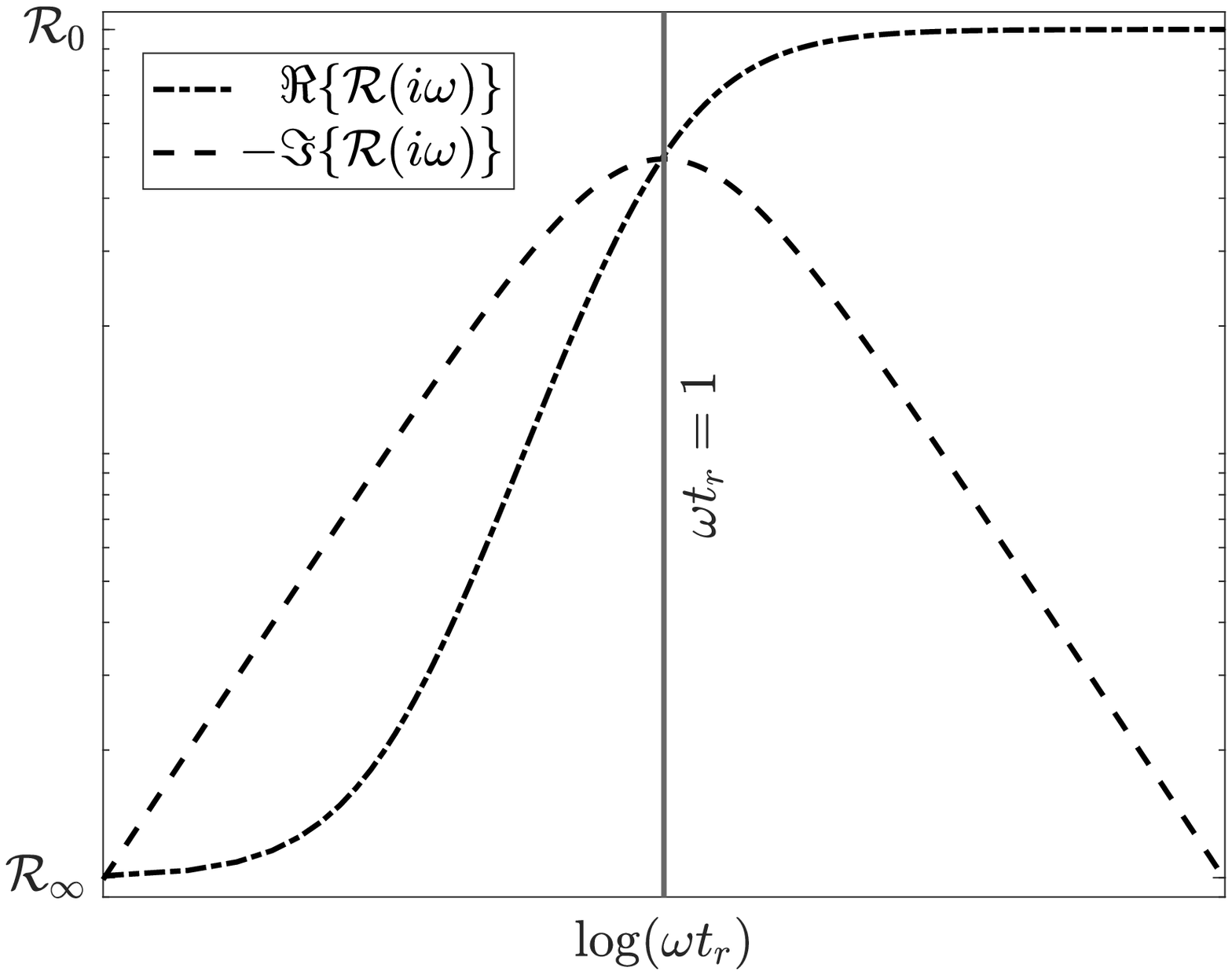}
\end{subfigure}
\caption{An example of prototype, single relaxation time, scalar relaxation function time-domain behaviour (left) given by (\ref{eqn:Modulus time domain}), and its frequency domain counterpart (right) from (\ref{eqn:Modulus freq domain}).}
\label{fig:SLM Rubber-Steel}
\vspace{-0.6cm}
\end{figure}

Stress relaxation tests aim to investigate the viscoelastic properties of a given sample of material via specific loading modes, e.g. shear, uniaxial or bi-axial compression, etc. A general expression for relaxation functions is the so-called \textit{Prony series} \cite{chen2000determining}, which takes the form 
\begin{align}
    \mathcal{R}(t) &= \left(\mathcal{R}_{\infty} + \sum_{n=1}^N \mathcal{R}_n e^{-t/t_n}\right)H(t),
\end{align}
where $H(\circ)$ denotes the Heaviside function and $t_n$ are characteristic relaxation times of the medium in question. $R_{\infty}$ is the associated \textit{long-term modulus}, resulting from the limit $t\rightarrow\infty$, whilst
$\mathcal{R}_0=\mathcal{R}_{\infty}+\sum_{n=1}^N\mathcal{R}_n$ is the \textit{instantaneous} modulus. In practice modes of deformation or propagation are chosen that can isolate the dependence of relaxation functions so that they can be measured experimentally \cite{balbi2018modified}. A common scenario for the purposes of modelling is to assume a single relaxation time ($t_1=t_r$), which is often referred to as the \textit{Standard linear solid model} (SLSM):
\begin{equation}\label{eqn:Modulus time domain}
    \mathcal{R}(t) = \left( \mathcal{R}_\infty +(\mathcal{R}_0-\mathcal{R}_\infty)\me^{-t/t_r}\right) H(t),
\end{equation}
where in practice the relaxation time is obtained by fitting the model to the relaxation test data \cite{chen2000determining}. Following (\ref{3complexRelaxationFuns}), (\ref{eqn:ComplexmRelaxation}), in the frequency domain (\ref{eqn:Modulus time domain}) becomes
\begin{equation}\label{eqn:Modulus freq domain}
    \mathcal{R}(\mi \omega) = \mathcal{R}_\infty - (\mathcal{R}_0 - \mathcal{R}_\infty)\frac{\mi \omega t_r }{1 - \mi \omega t_r},
\end{equation}
and it is apparent from (\ref{eqn:Modulus freq domain}) that, in the low frequency (long time, or \textit{rubbery}) and high frequency (short time, or \textit{glassy}) limits, $\mathcal{R}_\infty$ and $\mathcal{R}_0$ are respectively obtained (see Figure \ref{fig:SLM Rubber-Steel}). When separating (\ref{eqn:Modulus freq domain}) into real and imaginary parts, the "loss tangent" may be defined which is frequently used in order to characterize viscoelastic losses under steady state oscillatory conditions and associated experimental data  \cite{gottenberg1964experiment}. In practice, the ratio $\mathcal{R}_0/\mathcal{R}_\infty$ can be very large, up to several orders of magnitude, see e.g.\ \cite{kari2001dynamic} for the shear modulus of an unfilled crosslinked rubber material.

Temperature can play a very important role in the behaviour of the moduli \cite{tobolsky1952elastoviscous,jeong1987frequency}. Linear TVE theory allows only for dependence of the mechanical properties on the background temperature $T_0$ as is the case in an isothermal theory. Stress relaxation tests as described above are associated with specific modes of deformation and, therefore, the corresponding data obtained provides, e.g.\ the time-dependent Young's modulus (e.g. $\mathcal{R}(t)$ in (\ref{eqn:Modulus time domain})) under uniaxial compression or tension. On the other hand, several other experimental methods are used to approximate the shear modulus, e.g.\ \cite{jeong1987frequency}. As a result, one would expect that for an isotropic medium these 2 independent constants are sufficient to describe the continuum in consideration. It turns out that this is often not the case due to the required accuracy of the experiments, and tests involving primarily volumetric effects are necessary \cite{tschoegl2002poisson}. This is particularly evidenced for nearly incompressible elastic materials, and a method to determine ${\tilde{K}}(\mi \omega)$ was presented in \cite{lifshitz1965propagation}, where it is assumed that bulk loss is a constant fraction of the loss in shear. This assumption led to good agreement with the observed experimental results, for polyethylene (PE) and Plexiglass (PMMA) the bulk loss $(\operatorname{Im}\{{\tilde{K}}\}/\operatorname{Re}\{{\tilde{K}}\})$ represents 20\% of the shear loss, whereas in polystyrene the bulk loss was calculated to be around 0.1\%. Nevertheless, to this day, data for bulk losses in general materials remains difficult to find, as discussed in \cite{ivanova2010derivation}.

The frequency dependence of the specific heat and thermo-mechanical coupling term  in (\ref{eqn:ComplexzetaRelaxation}) and (\ref{eqn:ComplexmRelaxation}) are reported even less, and these quantities are usually considered static, although relaxation type phenomena of the specific heat has been observed, e.g. \cite{birge1987wide}. This discussion for VE behaviour together with the thermal properties illustrates the intricacies involved in the correct determination of many of the quantities appearing in a TVE model. As a result, in studies seeking more qualitative results over a wider range of materials, common simplifications are made. In \cite{kaliske1997formulation} it is argued that in most instances VE effects are mainly related to the isochoric part of the deformation and therefore if we write the Cauchy stress (\ref{eqn:timeharmonicstressTVE}) in terms of the isochoric and deviatoric parts we have (recalling (\ref{linearStresstensor decomposed}))
\begin{equation}
   {\hat{\bm{\sigma}}} = 2{\tilde{\mu}}(\mi {\omega})\hat{\bm{e}} + ({{K}} \tr(\bm{\varepsilon}) - {T}_0 \tilde{\mathcal{R}}_3(\mi {\omega})\hat{\theta})\bm{I},
\end{equation}
where ${K}$ becomes a real valued constant from which the value of ${\tilde{\lambda}}(\mi \omega)$ follows through (\ref{eqn:nonlocalTVE lambda}). In \cite{obaid2017understanding} it is instead assumed that the Young's modulus takes the form \eqref{eqn:Modulus freq domain}, whilst the Poisson's ratio is kept constant. In turn this implies that the shear modulus also takes the form \eqref{eqn:Modulus freq domain}. The magnitude of the variation in the specific heat is such that it will be assumed constant.

\subsubsection{Relaxation function interpretation of local TVE}\label{subsection:TVE->Local TVE}
The local TVE model discussed in Section \ref{sec:linearTVE}\ref{subsection:local TVE} can be thought of as a special case from that of Section \ref{sec:linearTVE}\ref{subsection: general TVE} where the kinematical and thermal time histories represented by ${\mathcal{R}}_1, {\mathcal{R}}_2, {\mathcal{R}}_3, {\mathcal{R}}_4$ in (\ref{TVE Non Local Eqns}) are described by Heaviside and delta functions. In the frequency domain, this simply results in the choice
\begin{equation}\label{eqns:localTVEquantities}
   { \tilde{\lambda}}(\mi \omega) = {\lambda} - \mi {\omega} {\eta}_\lambda, \quad {\tilde{\mu}}(\mi {\omega}) = {\mu} - \mi {\omega} {\eta}_\mu, \quad {\tilde{c}}_v(\mi {\omega}) = {c}_v, \quad {\tilde{\mathcal{R}}}_3(\mi {\omega}) = {\alpha} {K},
\end{equation}
in (\ref{eqn:timeharmonicstressTVE})-(\ref{eqn:timeharmonicmomentumTVE}) to arrive at the local TVE theory. In the time domain, the instantaneous local viscous effects are represented by delta functions (e.g. \cite{borcherdt2009viscoelastic}) such that for the shear modulus
\begin{equation}\label{eqn:time domain KV}
    {\mathcal{R}}_1(t) =2({\mu} H(t) + {\eta}_\mu \delta(t)).
\end{equation}
This can be deduced by taking the inverse Fourier transform of (\ref{eqn:ComplexshearRelaxation})${}_1$, and similarly for the bulk modulus. The time domain representation for the shear modulus (\ref{eqn:time domain KV}) shows how relaxation effects as discussed in Section \ref{sec:linearTVE}\ref{subsection: general TVE}\ref{subsection:Choice of moduli} are clearly not captured with local TVE. In the frequency domain the real part remains constant whereas the imaginary part becomes unbounded as the frequency increases. For this reason Local TVE is in general not suitable in studies beyond single frequency analyses. Given that in general we are interested in wave propagation in materials over rather general frequencies this is significantly  restrictive.

Next we consider asymptotic limits under which  thermo-compressional coupling can be significantly simplified in the context of the general TVE theory, before moving onto specific  physical limits in the next section.

\vspace{-0.35cm}
\subsection{Asymptotic approximations for thermo-compressional coupling}\label{subsection: asymptotics nonlocalTVE}

Here we simplify the thermo-compressional wavenumbers $ a\pm b = a \pm \sqrt{a^2 -k_\phi^2 k_\theta^2}$ in \eqref{eqn:DimTVEdecoupled wavenumbers kvarphi kvartheta}, and the temperature field~\eqref{TVE w relaxation temperature} by identifying one small parameter. We note that, different to several references stemming from~\cite{epstein1953absorption} for fluids, we find we only need one small parameter, rather than two,  to reach a simple and accurate model. Asymptotic analysis illustrates that $k_{\varphi}$ is a quasi-mechanical wavenumber, whilst $k_{\vartheta}$ is a quasi-thermal  wavenumber. Similar expressions for 1D TVE waves are given in \cite{christensen2012theory}, Section 6.3.

\begin{figure}
\centering
\begin{subfigure}{.43\textwidth}
  \centering
  \includegraphics[height=4cm]{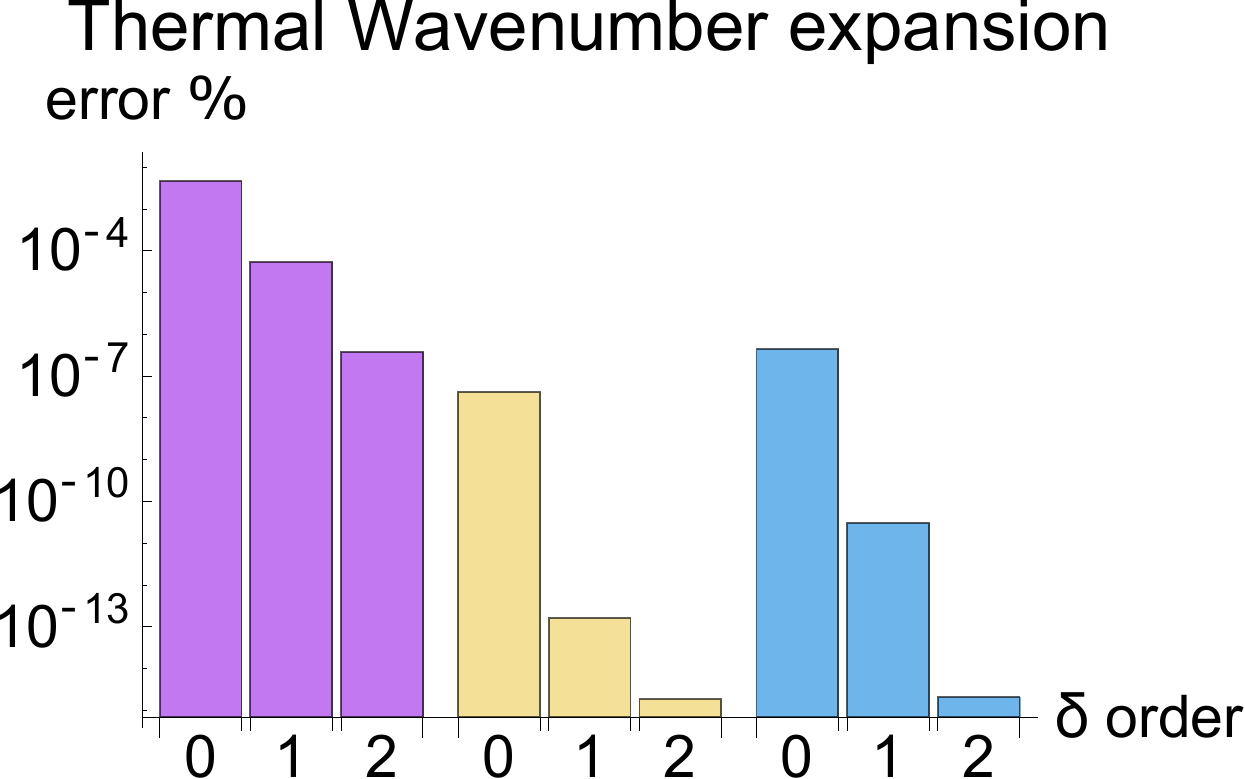}
  \end{subfigure}
\begin{subfigure}{.43\textwidth}
\centering
    \includegraphics[height=4cm]{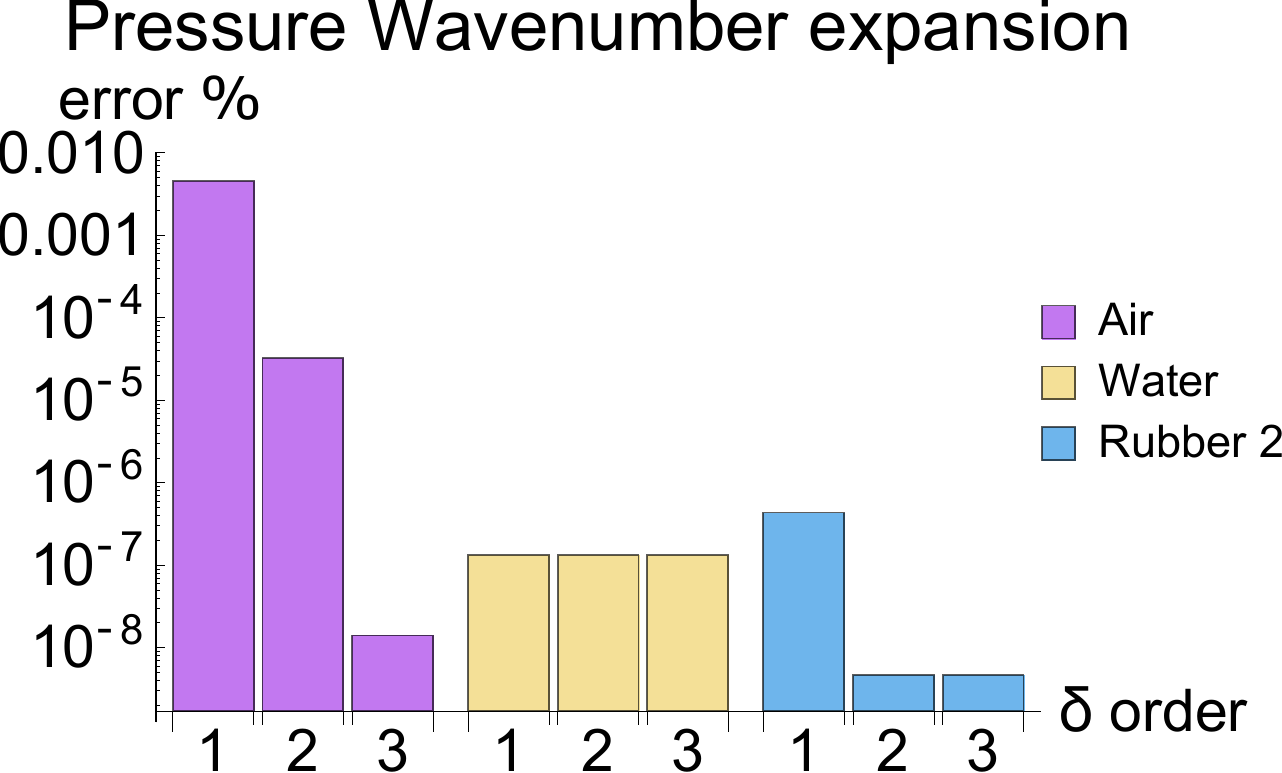}
\end{subfigure}
\caption{Maximum relative errors for the asymptotic expansions (\ref{asymptotics:Wavenumbers}) for a frequency range of $10$kHz to $10$MHz and material parameters from Table \ref{table: TVE parameters Rubber & Steel}. For Rubber 2, the shear modulus is described by the single Prony term relaxation function (\ref{eqn:Modulus freq domain}) where the frequencies cover both the rubber and glassy phase.}
\label{fig:Asymptotics}
\vspace{-0.6cm}
\end{figure}

For a vast range of frequencies and materials, including solids, liquids and gases, it can be observed that the pressure dominated wavelength is far longer than the thermal dominated wavelength, which leads us to the small parameter\footnote{Alternatively, we could just have assumed that $|{a}^2| \ll |{k}_\phi^2{k}_\theta^2|$, but this approach is avoided since its physical interpretation is not straightforward.}:
\begin{equation}\label{kphi << ktheta}
   |\delta| \ll 1 \quad \text{where} \quad 
   \delta = \frac{{k}_\phi^2}{ {k}_\theta^2} = 
  \frac{- \mi {\omega} {\mathscr{K}}}{{c}_v(\mi  \omega )({\tilde{\lambda}}(\mi {\omega}) + 2 {\tilde{\mu}}(\mi {\omega}))},  
\end{equation}
and we further assume that 
\begin{equation}\label{kthetasq/LthetaLphi}
     |\delta|  \ll   \left\lvert \frac{{L}_\phi L_\theta }{{k}_\theta^2} \right\rvert.
\end{equation}
 The right side of the inequality (\ref{kthetasq/LthetaLphi}) is a non-dimensional number related to the coupling between thermal and pressure modes. If the right side of (\ref{kthetasq/LthetaLphi}) is of the same order as $\delta$, or smaller, then the structure of the asymptotics below changes, as there will be  almost no coupling between thermal and pressure modes. To summarise, inequality (\ref{kthetasq/LthetaLphi}) is a necessary condition for these modes to be coupled. The inequality (\ref{kthetasq/LthetaLphi}) is also equivalent to ${\omega} {\mathscr{K}} \ll {c}_v {K} (\gamma - 1)$ which for a given a material can be a useful upper bound on the admissible frequency of the expansions below. Based on the discussion in Section \ref{sec:linearTVE}\ref{subsection: general TVE}\ref{subsection:Choice of moduli}, we will neglect thermal histories and thus write ${\tilde{c}}_v(\mi {\omega}) = {c}_v,$ ${\tilde{\mathcal{R}}}_3(\mi {\omega}) = {\alpha} {K}$.
Expanding in $\delta$ then, 
\begin{align*}
  {b} 
    & =  \pm \frac{1}{2} \left[{k}_\theta^2 -L_{\theta}{L}_{\phi} - \frac{{k}_\theta^4 + {k}_\theta^2 L_{\theta}{L}_{\phi} }{{k}_\theta^2 -L_{\theta}{L}_{\phi}} \delta 
     - \frac{ 2 {k}_\theta^6 L_{\theta}{L}_{\phi}}{({k}_\theta^2 -L_{\theta}{L}_{\phi})^3}\delta^2 + O(\delta^3)
    \right], 
\end{align*}
where the sign chosen depends on the complex argument of the term within the square-root, and the chosen branch cut. Depending on this choice we will have either $a \pm b =  k_\varphi^2$ and $a \mp b =  k_\vartheta^2$, with  $ k_\varphi^2$ and $ k_\vartheta^2$ shown below:
\begin{subequations}\label{asymptotics:Wavenumbers}
\vspace{-0.3cm}
\begin{align}
    {k}_\varphi^2 &=  \frac{{k}_\theta^4}{{k}_\theta^2 - L_\theta {L}_\phi} \delta + \frac{ {k}_\theta^6 L_\theta {L}_\phi}{({k}_\theta^2 - L_\theta {L}_\phi)^3} \delta^2 + O(\delta^3),
    \\
    {k}_\vartheta^2 &=  {k}_\theta^2 - L_\theta {L}_\phi - \frac{{k}_\theta^2 L_\theta {L}_\phi}{{k}_\theta^2 - L_\theta {L}_\phi} \delta + O(\delta^2).
\end{align}
\end{subequations}
Similarly, we can now expand the temperature contributions $( a \pm  b -  k_\phi^2)/L_\theta$ given in (\ref{temperature contributions}). 
We find
\begin{subequations}\label{temperature amplitudes approxs}
\begin{align}
  {\mathscr{T}}_\varphi &= \frac{1}{L_\theta}({k}_\varphi^2 -{k}_\phi^2) = \frac{ {k}_\theta^2 {L}_\phi}{{k}_\theta^2-L_\theta {L}_\phi} \delta +\frac{{k}_\theta^6 {L}_\phi}{\left({k}_\theta^2-L_\theta {L}_\phi\right)^3} \delta ^2 +  O(\delta^3),
  \\
    {\mathscr{T}}_\vartheta &= \frac{1}{L_\theta}( {k}_\vartheta^2 -{k}_\phi^2)  = \frac{{k}_\theta^2 - L_\theta {L}_\phi}{L_\theta} -
  \frac{ {k}_\theta^4}{ {k}_\theta^2-L_\theta {L}_\phi } \delta + O(\delta^2), 
\end{align}
\end{subequations}
so that ${\mathscr{T}}_\varphi$ ($ {\mathscr{T}}_\vartheta$) is the temperature contribution corresponding to the mode with wavenumber ${k}_\varphi$ (${k}_\vartheta$). An illustration of the accuracy of the expansions for the thermo-compressional wavenumbers for different materials is given in Figure \ref{fig:Asymptotics}. Similar results were obtained for the temperature contributions (\ref{temperature amplitudes approxs}) but have not been included here.

\section{Limits to theories that neglect specific physical effects} \label{sec:theorylimits}

A plethora of approximate thermo-visco-elastic theories exist that neglect certain physical effects. Here we describe such theories in terms of parameter limits of the general TVE theory described above, noting that we have already described how local TVE is recovered from non-local TVE in Section \ref{sec:linearTVE}\ref{subsection: general TVE}\ref{subsection:TVE->Local TVE} via the choice of specific relaxation functional forms.  More generally, it is important to understand how significant the neglected terms are when the full TVE is compared with the simpler theories. The efficacy of the various limits is thus studied with regard to a canonical problems involved half-spaces in Section \ref{sec:Application1:2HSs}.

\begin{figure}
\centering 
\def\svgwidth{\columnwidth}
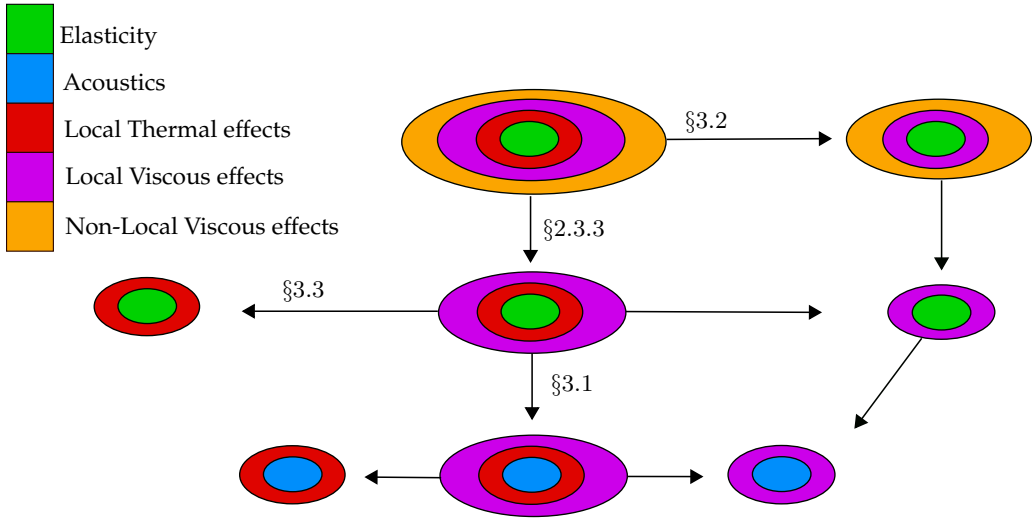
\caption{Representation of various elasto/acoustic dissipative theories, where the arrows indicate various limits that can be taken to arrive at other (more restrictive) theories.}
\label{Fig:Chapter3}
\vspace{-0.6cm}
\end{figure}

Figure \ref{Fig:Chapter3} summarises the various limits taken from the TVE theory in the frequency domain, starting from the current general framework, where various effects can be switched off and on to yield various commonly used theories. Other relevant dissipative theories concerning thermal relaxation in solids and those involving molecular relaxation effects in the acoustics of gases are not included since these require further modelling considerations, (see e.g.\ Section 2.4 in \cite{bruneau2013fundamentals}).

\subsection{Thermo-visco-acoustic (TVA) fluids}
\label{subsection: TVE-> TVA}
Starting with the local TVE theory described in Section \ref{sec:linearTVE}\ref{subsection:local TVE} and taking the standard limit of zero shear modulus, 
\begin{equation} \label{mu-> 0}
    {\mu} \rightarrow 0,
\end{equation}
leads to the widely used model for (local in time) thermo-visco-acoustics \cite{pierce1981acoustics,bruneau2013fundamentals}. 
In this regime, the thermodynamic identity (\ref{eqn:thermodIdentity}) becomes
\begin{equation}\label{eqn:thermidentTVAlimit 2}
    \gamma - 1 =  \frac{ {\alpha}^2 {T}_0 {c}_{\text{Iso}}^2}{{c}_v}, \qquad \text{where} \qquad {c}_{\text{Iso}}^2 = \frac{{\lambda}_{\text{Iso}}}{{\rho}_0},
\end{equation}
since in the limit $K_\text{Iso}\rightarrow \lambda_\text{Iso}$.
The subscript "Iso" in the definition of the isothermal sound speed ${c}_{\text{Iso}}$ is chosen to emphasize that these quantities are defined at a state of constant temperature\footnote{This distinction is often ignored for liquids and solids since it is not as important (see e.g. Section 1.9.2 of \cite{pierce1981acoustics}), but is paramount for gases.} (see e.g. (78) in \cite{ivanova2010derivation}). Note that here the Lam\'e parameters are isothermal by definition (${\lambda} \equiv {\lambda}_{\text{Iso}}$) since the Helmholtz free energy is expanded from a state of constant temperature  ${T}_0$ and zero strain (see (\ref{FreeEnergy1}) in Appendix \ref{local TVE derivation}). With (\ref{mu-> 0}) the thermo-mechanical coupling constant $L_\theta$ (\ref{eqn:lphi and ltau}) can be approximated by  $L_\theta \approx - {\alpha} {T}_0$
since for frequencies of interest we have ${\omega}{\eta}_\lambda, {\omega} {\eta}_\mu \ll {\lambda}$. Furthermore, with (\ref{mu-> 0}) and (\ref{eqn:thermidentTVAlimit 2}) the quantities  (\ref{eqn:TVEwavenumbers}) and (\ref{eqn:lphi and ltau}) become
\begin{subequations}\label{TVE qtys with mu->0}
\begin{align}
     {k}_{\theta}^2=\frac{\mi {\rho}_0 {\omega} {c}_p}{\gamma {\mathscr{K}}}, \label{DIMTVEkphi limit} \qquad 
     {k}_{\phi}^2 \rightarrow \frac{{\rho}_0 {\omega}^2}{{\rho}_0 {c}_{\text{Iso}}^2 - \mi {\omega} {\zeta}}, \qquad 
      {k}_{\Phi}^2 \rightarrow \frac{\mi {\rho}_0 {\omega}}{{\eta}_\mu},\\  \label{DIMTVE Lphi limit}
     {L}_\phi \rightarrow \frac{\mi  {\rho}_0 {\omega} {\alpha}  {c}_{\text{Iso}}^2}{{\mathscr{K}}}, \qquad  L_\theta \rightarrow -\frac{{\alpha} {\rho}_0 {c}_{\text{Iso}}^2 {T}_0}{{\rho}_0 {c}_{\text{Iso}}^2 - \mi {\omega} {\zeta}}, 
\end{align}
\end{subequations}
with ${\zeta} = {\eta}_\lambda + 2{\eta}_\mu$ and given that ${c}_v = {c}_p/\gamma $. With (\ref{TVE qtys with mu->0}) and in the limit $\mu \rightarrow 0$, the linear operator \eqref{eqn:order4} becomes
\begin{align}\label{tve operator tva limit 2}
{\mathcal{L}}_O \rightarrow {\mathcal{L}}_\text{TVA} = ({\rho}_0{c}_\text{A}^2 - \mi {\omega} {\zeta} \gamma) {\mathscr{K}} \mathcal{{\nabla}}^4 +\mi {\omega}[{\rho}_0^2 {c}_\text{A}^2{c}_p -\mi {\omega} {\rho}_0({c}_p {\zeta} + {\mathscr{K}} \gamma)] \mathcal{{\nabla}}^2 +\mi {\rho}_0^2 {c}_p {\omega}^3,
\end{align}
where we have made use of (\ref{eqn:thermidentTVAlimit 2}) in terms of the \textit{adiabatic} speed of sound ${c}_\text{A}$ as is common in acoustics with the relation ${c}_\text{A}^2 = \gamma {c}_\text{Iso}^2$. The operator (\ref{tve operator tva limit 2}) is identical to that in (2.70) of \cite{bruneau2013fundamentals} for TVA when the latter is written in the frequency domain and in the absence of any sources, which confirms that local TVE theory recovers TVA. Since the decomposition is unique up to a constant (as seen in (\ref{eqn:varphithetamatform})), in electronic supplementary material Section SM3 we explicitly match the current TVE potentials to those corresponding to TVA in \cite{cotterill2018thermo}.


\vspace{-0.5cm}
\subsection{Non-local (in time) visco-elasticity (VE)}
\label{subsection:TVE->VE}

Starting with the general non-local TVE theory described above and taking the limit of zero thermo-mechanical coupling\footnote{In the local TVE case we simply take the limit of zero thermal expansion coefficient, that is ${\alpha}T_0 \rightarrow 0$. }, that is
\begin{equation}\label{G3 relaxation goes to 0}
     {\tilde{\mathcal{R}}}_3(\mi {\omega}) \rightarrow 0,
\end{equation}
in (\ref{eqn:timeharmonicstressTVE}), (\ref{nonlocalTVE stress and energy with relaxation}), results in ${\hat{\bm{\sigma}}}^{\text{VE}} = 2{\tilde{\mu}}(\mi {\omega}) \hat{\bm{\varepsilon}} + {\tilde{\lambda}}( \mi {\omega}) \tr(\hat{\bm{\varepsilon}})\bm{I}$, 
as well as
\begin{subequations}\label{VE eqnsss}
\begin{align} \label{eqn:timeharmonicenergyVEE}
     \left( \Delta  + \frac{\mi {\omega} {\rho}_0  {\tilde{c}}_v(\mi {\omega})}{{\mathscr{K}}} \right) \hat{\theta}^{\text{VE}} &= 0,\\ \label{eqn:timeharmonicmomentumVEE}
    ({\tilde{\lambda}}( \mi {\omega}) + 2{\tilde{\mu}}(\mi {\omega})) {\mathbf{\nabla}} \left({\mathbf{\nabla}} \cdot {\hat{\mathbf{u}}} \right) - {\tilde{\mu}}(\mi {\omega}) {\mathbf{\nabla}} \times {\mathbf{\nabla}} \times  {\hat{\mathbf{u}}} +{\rho}_0 {\omega}^2 {\hat{\mathbf{u}}} &= \mathbf{0}.
\end{align}
\end{subequations}
These are the governing equations for visco-elasticity, including stress relaxation. It is apparent in (\ref{VE eqnsss}) that there is no longer coupling between kinematic and thermal effects, and hence the wave potentials directly give 
\begin{align}
    {\hat{\mathbf{u}}}^\text{VE} = {\nabla} {\phi} + {\nabla} \times {\mathbf{\Phi}}, \label{viscoelastic helmholtz decomposition}
\end{align}
where
\begin{subequations} \label{viscoelastic phi&Phi eqn}
\begin{align} 
    &\left(\Delta  + \frac{{\rho}_0 {\omega}^2}{{\tilde{\lambda}}(\mi {\omega}) + 2 {\tilde{\mu}}(\mi {\omega})}\right){\phi} = 0,  
    & \left(\Delta  + \frac{{\rho}_0 {\omega}^2}{{\tilde{\mu}}(\mi {\omega})}\right){\mathbf{\Phi}} = \mathbf{0},
\end{align}
\end{subequations}
recalling that the Lam\'e parameters in (\ref{viscoelastic phi&Phi eqn}) are isothermal\footnote{For the particular relations with the corresponding adiabatic moduli, see e.g. \cite{lubarda2004thermodynamic}.}. Nevertheless, in practice it is important to understand the effect of this limit on the decomposition that leads to the corresponding TVE wave potentials. It is clear that the shear wave potential remains unchanged in the limit (since it is independent of thermal effects). The situation for the thermo-compressional fields is slightly more subtle. Direct substitution of (\ref{G3 relaxation goes to 0}) into (\ref{eqn:decoupledCURLpart}), (\ref{eqn:varthetahelmholtz}) and (\ref{eqn:varphihelmholtz}) with \eqref{eqns:kdiff} leads to
\begin{equation}\label{VE limit potentials}
    {\varphi} \rightarrow C_1{\phi}, \quad {\vartheta} \rightarrow {C}_2 \theta^{\text{VE}}, \quad \text{since} \quad {a}-{b} \rightarrow  \frac{{\rho}_0 {\omega}^2}{{\tilde{\lambda}}(\mi {\omega}) + 2 {\tilde{\mu}}(\mi {\omega})}={k}_\phi^2, \quad {a}+{b} \rightarrow \frac{\mi {\omega} {\rho}_0  {\tilde{c}}_v(\mi {\omega})}{{\mathscr{K}}}={k}_\theta^2,
\end{equation}
for some constants $C_1, {C}_2$ arising due to the uniqueness of the linear PDE solution being up to a constant. However, direct comparison between the curl free components of the TVE and VE displacements (\ref{TVE w relaxation displacement}), (\ref{viscoelastic helmholtz decomposition}) implies $C_1 = 1$ and  ${\vartheta} \rightarrow 0$ which restricts the form of ${C}_2$ but this is not sufficient to determine it explicitly. Therefore, in order to find this constant we consider the effect of the limit $(\ref{G3 relaxation goes to 0})$ on the TVE temperature (\ref{TVE w relaxation temperature}). We obtain ${\mathscr{T}}_\varphi \rightarrow 0$, which gives $\theta^{\text{TVE}} \rightarrow ({k}_\theta^2 - {k}_\phi^2) {C}_2 \theta^{\text{VE}}/L_\theta$ 
(after using the second equation of (\ref{VE limit potentials})) so that in order to recover the VE solution we must choose
\begin{equation}
    {C}_2 = \frac{L_\theta}{{k}_\theta^2 - {k}_\phi^2} = \frac{\mi {T}_0 {\tilde{\mathcal{R}}}_3(\mi {\omega}) {\mathscr{K}}}{{\rho}_0 {\omega}({c}_v(\mi {\omega})({\tilde{\lambda}}(\mi {\omega})+2{\tilde{\mu}}(\mi {\omega}))+\mi {\omega} {\mathscr{K}}},
\end{equation}
from which it is clear that in the limit both $\theta^{\text{TVE}} \rightarrow  \theta^{\text{VE}}$ and ${\vartheta} \rightarrow 0$ as required. Furthermore, (local) visco--acoustic Newtonian fluids (e.g. \cite{scharstein2007acoustic}) can also be described by (\ref{VE eqnsss})-(\ref{viscoelastic phi&Phi eqn}) by further letting $\mu \rightarrow 0$ so that ${\mu}(\omega) = -\mi \omega \eta_\mu$ which is a convenient way to model viscous fluids like water \cite{wu1995alternative, cegla2005material}.

\vspace{-0.35cm}
\subsection{Thermo-elasticity (TE)}

The final simplified theory is the case when viscous dissipation is neglected, leading to the theory of linear thermo-elasticity. In the frequency domain this can be thought of as the local TVE model presented in Section \ref{sec:linearTVE}\ref{subsection:local TVE} with real-valued Lam\'e parameters. Indeed let 
\begin{equation}\label{eqns:thermoelasticquantities}
   { \tilde{\lambda}}(\mi \omega) = {\lambda}, \quad {\tilde{\mu}}(\mi {\omega}) = {\mu}, \quad {\tilde{c}}_v(\mi {\omega}) = {c}_v, \quad {\tilde{\mathcal{R}}}_3(\mi {\omega}) = {\alpha} {K}, 
\end{equation}
and substitute (\ref{eqns:thermoelasticquantities}) in (\ref{eqn:timeharmonicstressTVE}) and (\ref{nonlocalTVE stress and energy with relaxation}) so that we obtain the Cauchy stress ${\hat{\bm{\sigma}}}^{\text{TE}} = 2 {\mu} \hat{\bm{\varepsilon}} + ({\lambda} \tr(\hat{\bm{\varepsilon}}) -  {\alpha} {K} {T}_0 \hat{\theta})\bm{I},$ and the corresponding equations for time-harmonic thermo-elasticity \cite{boley2012theory}
\begin{subequations}\label{ThermoElasticity eqns}
\begin{align}
    \label{eqn:timeharmonicenergyTE}
    &{\mathscr{K}} \Delta \hat{\theta} + \mi {\omega} {\rho}_0 {c}_v \hat{\theta} +  \mi {\omega} {\alpha} {K}  {\mathbf{\nabla}} \cdot {\hat{\mathbf{u}}} = 0,\\ \label{eqn:timeharmonicmomentumTE}
   & ({{\lambda}} + 2{\mu}) {\mathbf{\nabla}} \left({\mathbf{\nabla}} \cdot {\hat{\mathbf{u}}} \right) - {{\mu}} {\mathbf{\nabla}} \times {\mathbf{\nabla}} \times  {\hat{\mathbf{u}}} - {\alpha} {K} {T}_0  {\mathcal{\nabla}} \hat{\theta} +{\rho}_0 {\omega}^2 {\hat{\mathbf{u}}} = \mathbf{0}.
\end{align}
\end{subequations}
The structure of (\ref{eqn:timeharmonicenergyTE}), (\ref{eqn:timeharmonicmomentumTE}) allows for the same decomposition $ {\hat{\mathbf{u}}}^{\text{TE}} =  {\nabla}( {\vartheta} + {\varphi}) +  {\nabla}  \times {\mathbf{\Phi}}$,
where the wave potentials must still satisfy (\ref{eqn:decoupledCURLpart}), (\ref{eqn:varthetahelmholtz}) and (\ref{eqn:varphihelmholtz}) with simplified TVE parameters in (\ref{eqn:TVEwavenumbers}), (\ref{eqn:lphi and ltau}) becoming real valued and frequency independent, i.e.
\begin{align}\label{DIMTE: general elements of wavenumbers}
    {k}_{\phi}^2=\frac{{\rho}_0 {\omega}^2}{{\lambda}+2{\mu}}, \quad {k}_{\Phi}^2=\frac{{\rho}_0 {\omega}^2}{{\mu}}, \quad L_\theta = - \frac{{T}_0 {\alpha}{K}}{{\lambda}+2{\mu}},
\end{align}
whereas ${k}_\theta^2$, ${L}_\phi$ remain unchanged.

\section{Two TVE half spaces in perfect contact}\label{sec:Application1:2HSs}

To put our framework into practice, we next consider a forced boundary value problem (BVP) consisting of two TVE half-spaces. In the absence of thermal effects (using the theory presented in Section \ref{sec:theorylimits}\ref{subsection:TVE->VE}) a detailed analysis for this problem is given in \cite{borcherdt2009viscoelastic}, who generalized the work pioneered by \cite{becker1970ultrasonic} to include attenuation in reflection/transmission problems for ultrasonics. More recent work has included the presence of voids \cite{tomar2014time} or thermal relaxation \cite{das2020reflection}, but only for a single traction free half-space, presumably because their goal was to understand loss mechanisms for solids.

Here we are interested in interactions between different TVE media when in contact and in particular those that are deemed as "fluid" and "solid". With two half spaces we can illustrate the advantages of the general TVE model, the limits discussed in Section \ref{sec:theorylimits}, as well as the importance of stress relaxation effects (non-local in time) presented in Section \ref{sec:linearTVE}\ref{subsection: general TVE} as opposed to the local TVE version in Section \ref{sec:linearTVE}\ref{subsection:local TVE}, which is a common "go to" theory when experiments are performed at specific frequencies. 
\vspace{-0.3cm}

\subsection{Problem formulation}

\begin{figure}
\centering 
\def\svgwidth{\columnwidth}
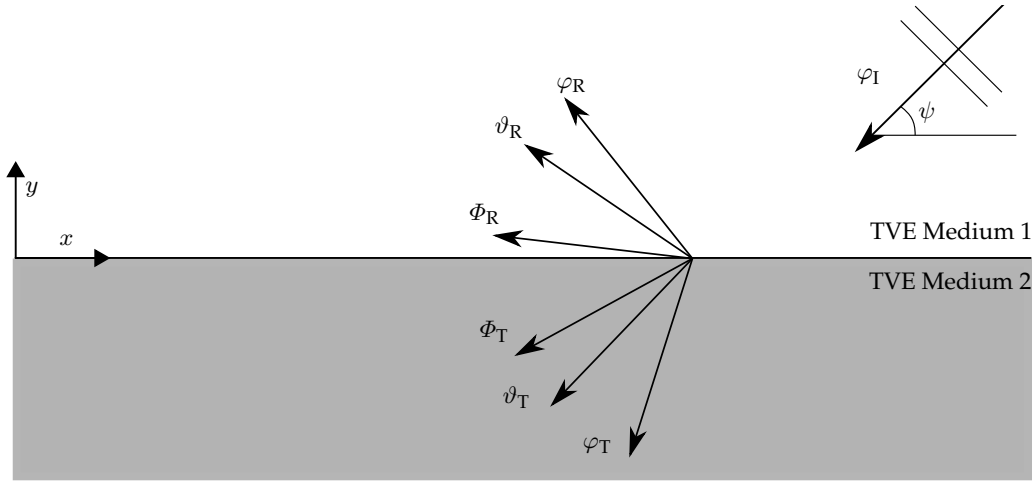
\caption{Schematic representation of the configuration of the problem of two welded semi-infinite TVE media. An incident P-dominated bulk mode impinging on the interface of the two distinct TVE domains gives rise to three reflected modes and three transmitted modes.}
\label{Fig:TVE1vsTVE2}
\vspace{-.77cm}
\end{figure}

We consider a plane-strain problem consisting of two distinct TVE half spaces in perfect contact at an interface along $y=0$, see Figure \ref{Fig:TVE1vsTVE2}. All of the quantities have been non-dimensionalised following Appendix \ref{appendix:non-dimensionalization}, and relevant dimensional parameters are distinguished by an overbar.

We choose the forcing to be a pressure-dominated plane wave
\begin{equation}\label{TVE-TVE incident P wave}
    \varphi_\text{I} = \me^{- \mi k_{\varphi_1}(x \cos \psi + y \sin \psi)}, \qquad \psi \in (0, \pi),
\end{equation}
where $\psi$ is the angle of incidence  (measured anticlockwise from $y=0$), and we assume $\operatorname{Re} k_{\varphi_1 } \geq 0$ and $\operatorname{Im}{k_{\varphi_1 }} \geq 0$. This incoming energy will be converted into reflected/transmitted thermo-compressional and shear modes. Given the translational invariance of the problem in the $x$-direction, each potential will depend on $x$ through $\me^{- \mi k_{\varphi_1}x \cos \psi}$, and therefore we write
\begin{subequations}\label{TVE HS R/T amplitudes}
\begin{align} \label{TVE HS P wave reflected amplitudes}
\varphi_\text{R} &= R_{\varphi } \me^{ \mi k_{\varphi_1} \sin \psi y}\me^{- \mi k_{\varphi_1}x \cos \psi},  
& \varphi_\text{T} &= T_{\varphi } \me^{- \mi d_{\varphi_\text{T}} y}\me^{- \mi k_{\varphi_1}x \cos \psi},\\ 
\label{TVE HS Th wave reflected amplitudes}
\vartheta_\text{R} &= R_{\vartheta} \me^{ \mi d_{\vartheta_\text{R}} y} \me^{- \mi k_{\varphi_1}x \cos \psi},  &
\vartheta_\text{T} &= T_{\vartheta} \me^{- \mi d_{\vartheta_\text{T}} y} \me^{- \mi k_{\varphi_1}x \cos \psi}, \\ \label{TVE HS SV wave reflected amplitudes}
    \mathbf \Phi_\text{R} & = \mathbf{e_z} R_{\Phi} \me^{ \mi d_{\Phi_\text{R}} y} \me^{- \mi k_{\varphi_1}x \cos \psi},  &
    \mathbf \Phi_\text{T} & = \mathbf{e_z} T_{\Phi} \me^{- \mi d_{\Phi_\text{T}} y}\me^{- \mi k_{\varphi_1}x \cos \psi},
\end{align}
\end{subequations}
where the potentials $\varphi_\text{R}$, $\vartheta_\text{R}$, and $\Phi_\text{R}$ are defined in the upper half space $y \geq 0$, while the potentials $\varphi_\text{T}$, $\vartheta_\text{T}$, and $\Phi_\text{T}$ are defined in the lower half space $y\leq 0$. We use subscripts R/T to denote reflected/transmitted respectively. In the above we introduced the notation
\begin{align*}\label{definining d's}
    d_{\varphi_\text{T}} &=  \mi \sqrt{-(k^2_{\varphi_2} - k^2_{\varphi_1} \cos^2 \psi)}, \quad 
    d_{\vartheta_\text{R}} = \mi \sqrt{-(k^2_{\vartheta_1} - k^2_{\varphi_1} \cos^2 \psi)}, \quad
    d_{\vartheta_\text{T}} = \mi \sqrt{-(k^2_{\vartheta_2} - k^2_{\varphi_1} \cos^2 \psi)}, 
    \\
    d_{\Phi_\text{R}} &= \mi \sqrt{-(k^2_{\Phi_1} - k^2_{\varphi_1} \cos^2 \psi)}, \quad d_{\Phi_\text{T}} = \mi \sqrt{-(k^2_{\Phi_2} - k^2_{\varphi_1} \cos^2 \psi)}.\numberthis
\end{align*}
With (\ref{definining d's}), when using the standard branch cut for the square root along the negative real line we have
\begin{equation}\label{Condition on imaginary parts}
    \operatorname{Im} d_{\varphi_\text{T}}, \operatorname{Im}d_{\vartheta_\text{R}}, 
    \operatorname{Im} d_{\vartheta_\text{T}}, \operatorname{Im}d_{\Phi_\text{R}},
    \operatorname{Im}d_{\Phi_\text{T}} 
    \geq 0,
\end{equation} 
which guarantees that each of the potentials in \eqref{TVE HS R/T amplitudes} are bounded within their respective half spaces. In order to completely determine the potentials~\eqref{TVE HS R/T amplitudes} we use the boundary conditions representing continuity of traction\footnote{Where in component form we have $\hat{\boldsymbol \sigma}_1  \mathbf{e_y} = ((\hat{\boldsymbol \sigma}_1)_{xy},(\hat{\boldsymbol \sigma}_1)_{yy},(\hat{\boldsymbol \sigma}_1)_{zy})$.}, displacement, temperature, and temperature flux
\begin{subequations}\label{TVE 2 HS BCs 1}
\begin{align}\label{TVE1vsTVE2 BC1}
    &\hat{\boldsymbol \sigma}_1  \mathbf{e_y} = \hat{\boldsymbol \sigma}_2 \mathbf{e_y}, \;\;
    \hat{\mathbf{u}}_1 = \hat{\mathbf{u}}_2
    \\
    &\hat{\theta}_1 = \hat{\theta}_2, \;\;
    \mathscr{K}_1 \nabla \hat{\theta}_1 \cdot \mathbf{e_y} = \mathscr{K}_2 \nabla \hat{\theta}_2 \cdot \mathbf{e_y} , \label{TVE1vsTVE2 BC2}
\end{align}
\end{subequations}
across $y =0$, where $\hat{\boldsymbol \sigma}_1$ and $\hat{\boldsymbol \sigma}_2$ are the stress tensors in the upper (1) and lower (2) media respectively, while $\mathbf u_1$ and $\mathbf u_2$ are the displacements in media 1 and 2.

Substituting (\ref{TVE HS R/T amplitudes}) into (\ref{TVE 2 HS BCs 1}),  using (\ref{eqn:timeharmonicstressTVE}) and (\ref{eqns:TVE w relaxation}), leads to the following six equations 
\begin{align}\label{TVE1vsTVE2 Scat system}
\begin{pmatrix}
a_{11} & a_{12} & a_{13} & a_{14} & a_{15} & a_{16} \\
a_{21} & a_{22} & a_{23} & a_{24} & a_{25} & a_{26} \\
a_{31} & a_{32} & a_{33} & a_{34} & a_{35} & a_{36} \\
a_{41} & a_{42} & a_{43} & a_{44} & a_{45} & a_{46} \\
a_{51} & a_{52} & a_{53} & a_{54} & 0 & 0 \\
a_{61} & a_{62} & a_{63} & a_{64} & 0 & 0 
\end{pmatrix} \begin{pmatrix} R_\varphi \\
T_\varphi \\
R_\vartheta \\
T_\vartheta \\
R_\Phi\\
T_\Phi \\
\end{pmatrix}  =\begin{pmatrix} 
\phantom{-}a_{11} \\
-a_{21} \\
-a_{31} \\
\phantom{-}a_{41} \\
-a_{51}\\
\phantom{-}a_{61} 
\end{pmatrix}.
\end{align}
To calculate the entries $a_{ij}$ we provide a Mathematica notebook as supplementary material \cite{artGitHub}. The above can be used to uniquely determine the six amplitudes $R_\varphi, T_\varphi, R_\vartheta, T_\vartheta, R_\Phi,$ and $T_\Phi$. 

\vspace{-0.3cm}
\subsection{The VE-VE limit}
In the limit of no thermal coupling we let $\alpha_1$, $\alpha_2 \rightarrow 0$ (and hence $k_{\varphi_1}\rightarrow k_{\phi_1}$, $k_{\varphi_2}\rightarrow k_{\phi_2}$ and $\mathscr{T}_{\varphi_1}$, $\mathscr{T}_{\varphi_2} \rightarrow 0$) in (\ref{TVE1vsTVE2 Scat system}) as discussed in Section \ref{sec:theorylimits}\ref{subsection:TVE->VE}. From this we conclude that $R_\vartheta, T_\vartheta \rightarrow 0$ and the scattering system reduces to
\begin{align}\label{VE1vsVE2 Scat system}
\begin{pmatrix}
a_{11} & a_{12}  & a_{15} & a_{16} \\
a_{21} & a_{22}  & a_{25} & a_{26} \\
a_{31} & a_{32}  & a_{35} & a_{36} \\
a_{41} & a_{42}  & a_{45} & a_{46} \\
\end{pmatrix} \begin{pmatrix} R_\varphi \\
T_\varphi \\
R_\Phi\\
T_\Phi \\
\end{pmatrix}  =\begin{pmatrix} 
\phantom{-}a_{11} \\
-a_{21} \\
-a_{31}\\
\phantom{-}a_{41} 
\end{pmatrix},
\end{align}
where the limit of $\alpha_1,\alpha_2 \to 0$ should be taken for each of the $a_{ij}$. For normal incidence, $\psi = \pi/2$ we obtain the classical solutions
\begin{subequations}\label{VE normal incident solutions}
\begin{align}
    R_\varphi=& \frac{-k_{\phi_1}(\tilde{\lambda}_1 + 2 \tilde{\mu}_1) + k_{\phi_2}(\tilde{\lambda}_2 + 2 \tilde{\mu}_2)}{k_{\phi_1}(\tilde{\lambda}_1 + 2 \tilde{\mu}_1) + k_{\phi_2}(\tilde{\lambda}_2 + 2 \tilde{\mu}_2)}=\frac{\bar{\rho}_2 \bar{c}_{\phi_2}-\bar{\rho}_1 \bar{c}_{\phi_1}}{\bar{\rho}_2 \bar{c}_{\phi_2}+\bar{\rho}_1 \bar{c}_{\phi_1}} , 
    \\
    T_\varphi=& \frac{2k_{\phi_1}^2(\tilde{\lambda}_1 + 2 \tilde{\mu}_1)}{k_{\phi_2}[k_{\phi_1}(\tilde{\lambda}_1 + 2 \tilde{\mu}_1) + k_{\phi_2}(\tilde{\lambda}_2 + 2 \tilde{\mu}_2)]}=\frac{2 \bar{\rho}_1 \bar{c}_{\phi_2}}{\bar{\rho}_2 \bar{c}_{\phi_2}+\bar{\rho}_1 \bar{c}_{\phi_1}},
\end{align}
\end{subequations}
where we have introduced the free space compressional wave speed in each medium through the relation $\bar{c}_\phi =  \bar{\omega}/\bar{k}_\phi$. The well-known equations (\ref{VE normal incident solutions}) give a clear interpretation of the role of the mechanical impedance $\Bar{\rho} \Bar{c}_\phi$ when it comes to reflection/transmission, see e.g. \S 1.4. in \cite{achenbach2012wave} (for elasticity). We next discuss the more subtle aspect of the partition of energy at the interface.


\subsection{Energy partitioning at the interface}

Consider the energy flux through the boundary $y=0$. The average energy flux vectors are defined in \eqref{eqn: time average energy flux}, and since for this problem we have two distinct media, we write\footnote{Where the product between the Cauchy stress and velocity is written in component form as $\bm{\sigma}_1  \dot{\mathbf{u}}_1^{*}={(\bm{\sigma}_1)}_{ij}(\dot{\mathbf{u}}_1)_{j}^{*}$ where we sum over $j$.}
\begin{equation}
 \langle  \mathbf{J} \rangle = \begin{cases}
    \langle  \mathbf{J}_1 \rangle = -\frac{1}{2} \operatorname{Re} \{{\bm{\sigma}}_1 \dot{{\mathbf{u}}}_1^{*}  + {\theta}_1 \mathscr{K}_1 \nabla {\theta}_1^* \}  \quad \text{for} \quad  y \geq 0, \\
     \langle  \mathbf{J}_2 \rangle = -\frac{1}{2} \operatorname{Re} \{{\bm{\sigma}}_{2} \dot{{\mathbf{u}}}_2^{*}  + {\theta}_2 \mathscr{K}_2 \nabla {\theta}_2^*\}  \quad \text{for} \quad  y < 0. 
     \end{cases}
\end{equation}
If the boundary conditions (\ref{TVE 2 HS BCs 1}) have been correctly enforced, we expect to have
\begin{equation}\label{eqn: continuity of normal energy flux}
  \langle  \mathbf{J}_1
  \rangle \cdot \mathbf{e_y}=  \langle  \mathbf{J}_2 \rangle \cdot \mathbf{e_y} \quad \text{at} \quad  y = 0,
\end{equation}
meaning that the normal component of the mean energy flux (or power per unit area averaged over a period) is \textit{continuous} across the boundary $y=0$. It is shown in electronic supplementary material Section SM4 how in order to exploit the role of each mode (\ref{eqn: continuity of normal energy flux}) can be written in terms of \textit{energy ratios} for reflected, transmitted and interacting modes with respect to the incident mode, which we write as
\begin{equation}\label{conservation of energy}
E_{R_\varphi}+E_{R_\vartheta} +E_{R_\Phi}+
E_{IR_\text{IR}}+
E_{IR_\text{RR}}+
E_{T_\varphi}+E_{T_\vartheta} +E_{T_\Phi}+E_{IT_\text{TT}}=1.
\end{equation}
After solving for all the relevant wave potentials, the above can be used as a check to ensure both numerical accuracy and algebraic correctness. We have noted that the presence of `crossed terms' (electronic supplementary material Section SM4) represented by interaction coefficients $E_{IR}, E_{IT}$ in (\ref{conservation of energy}) has been repeatedly ignored in the literature without justification e.g. \cite{das2020reflection,tomar2014time}. We find (not shown) that despite their contribution being small at lower frequencies, their importance in the energy balance equation becomes essential at higher frequencies, and it should therefore be emphasized under what conditions it is a valid approximation to ignore them. Further details can be found in \cite{borcherdt2009viscoelastic} (in the absence of thermal coupling).

\subsection{Numerical results and discussion}\label{subsection:Results and discussion}

We now present some illustrations of  
numerical solutions of the general system (\ref{TVE1vsTVE2 Scat system}) for specific pairs of TVE materials. All results were checked to accurately satisfy the energy flux balance (\ref{conservation of energy}). We thus demonstrate when thermal or viscous effects are important for these examples, and in particular we can illustrate the effect of stress relaxation. We do this by comparing solutions from the general TVE-TVE case in (\ref{TVE1vsTVE2 Scat system}) with the solutions of VE-VE (\ref{VE1vsVE2 Scat system}), which ignores thermal effects, the TVA-rigid solutions (\ref{TVE1vsRigid SolutionsISO}), which consider no transmission, and other variations specified in Table \ref{table: Acronym}.

We use typical values for air, water, steel and rubber as summarized in Table \ref{table: TVE parameters Rubber & Steel}. The large parameter space involved allows for an incredibly wide range of materials to be considered. Here we only consider a small fraction of this space, but hope that this work enables further exploration in the future. In particular we stress that the general TVE framework allows general materials to be considered and no distinction to be required between fluids or solids, etc.\ which frequently hampers progress via the necessary use of distinct notation for each medium.

\begin{table}
\centering
\small
\begin{tabular}{ |m{1.6cm}||m{1.4cm}|m{1.4cm}|m{1cm}|m{1cm}|m{1.05cm}|m{1cm}|m{1cm}|  }
 \hline
 Acronym &   TVA--Local TVE & VA--Local VE & TVA--Rigid & A--Rigid &TVE--TVA  & VE--VA   & TVA--TVE \\ \hline
 Equation &   (\ref{TVE1vsTVE2 Scat system}) with (\ref{moduli TVA-LocalTVE})& (\ref{VE1vsVE2 Scat system}) with (\ref{moduli TVA-LocalTVE}) & (\ref{TVE Halfspace Fields ISO ALL}) with (\ref{TVE1vsRigid SolutionsISO}) & (\ref{TVE Halfspace Fields ISO ALL}) with (\ref{A1vsRigid SolutionsISO}) & (\ref{TVE1vsTVE2 Scat system}) with (\ref{moduli TVE-TVA})& (\ref{VE1vsVE2 Scat system}) with (\ref{moduli TVE-TVA}) & (\ref{TVE1vsTVE2 Scat system}) with (\ref{moduli TVA-TVE})\\
 \hline
\end{tabular}
\caption{Specific equations corresponding to the various acronyms used in the results and discussion of Section \ref{sec:Application1:2HSs}\ref{subsection:Results and discussion}.}
\label{table: Acronym}
\vspace{-0.6cm}
\end{table}

\subsubsection{TVA-Local TVE: Thermo-visco-elastic effects and fluid-structure interaction (FSI)}\label{results:TVA-LocalTVE}

 In this first instance we restrict the material parameters of medium 1 to those of air/water whereas for medium 2 we will concentrate on steel/rubber. Both air and water have many applications, while investigating steel and rubber means we are considering both soft and hard solids. We first investigate the use of TVA in medium 1 (or local TVE with  $\mu_1 = 0$) and local TVE in medium 2 such that the complex moduli appearing in (\ref{TVE1vsTVE2 Scat system}) are given by
 \begin{equation}\label{moduli TVA-LocalTVE}
     \tilde{\mu}_1(\mi \omega) =  - \mi \omega \eta_{\mu_1}, \quad \tilde{\lambda}_1(\mi \omega) = \lambda_1 - \mi \omega \eta_{\lambda_1}, \quad \tilde{\mu}_2(\mi \omega) =  \mu_2 - \mi \omega \eta_{\mu_2}, \quad \tilde{\lambda}_2(\mi \omega) = \lambda_2 - \mi \omega \eta_{\lambda_2},
 \end{equation}
as discussed in Sections \ref{sec:linearTVE}\ref{subsection: general TVE}\ref{subsection:TVE->Local TVE}, \ref{sec:theorylimits}\ref{subsection: TVE-> TVA}. For some parameters, it is difficult to find numerical values in the literature, take for example $\eta_{\lambda_2}$  \cite{ivanova2010derivation}. In these cases, we attempt to use reasonable values based on similar materials.  The viscoelastic parameters for steel are taken from Table 6.2.2. in \cite{borcherdt2009viscoelastic}. \\

\begin{figure}
  \begin{subfigure}[t]{.45\textwidth}
    \centering
    \includegraphics[width=\linewidth]{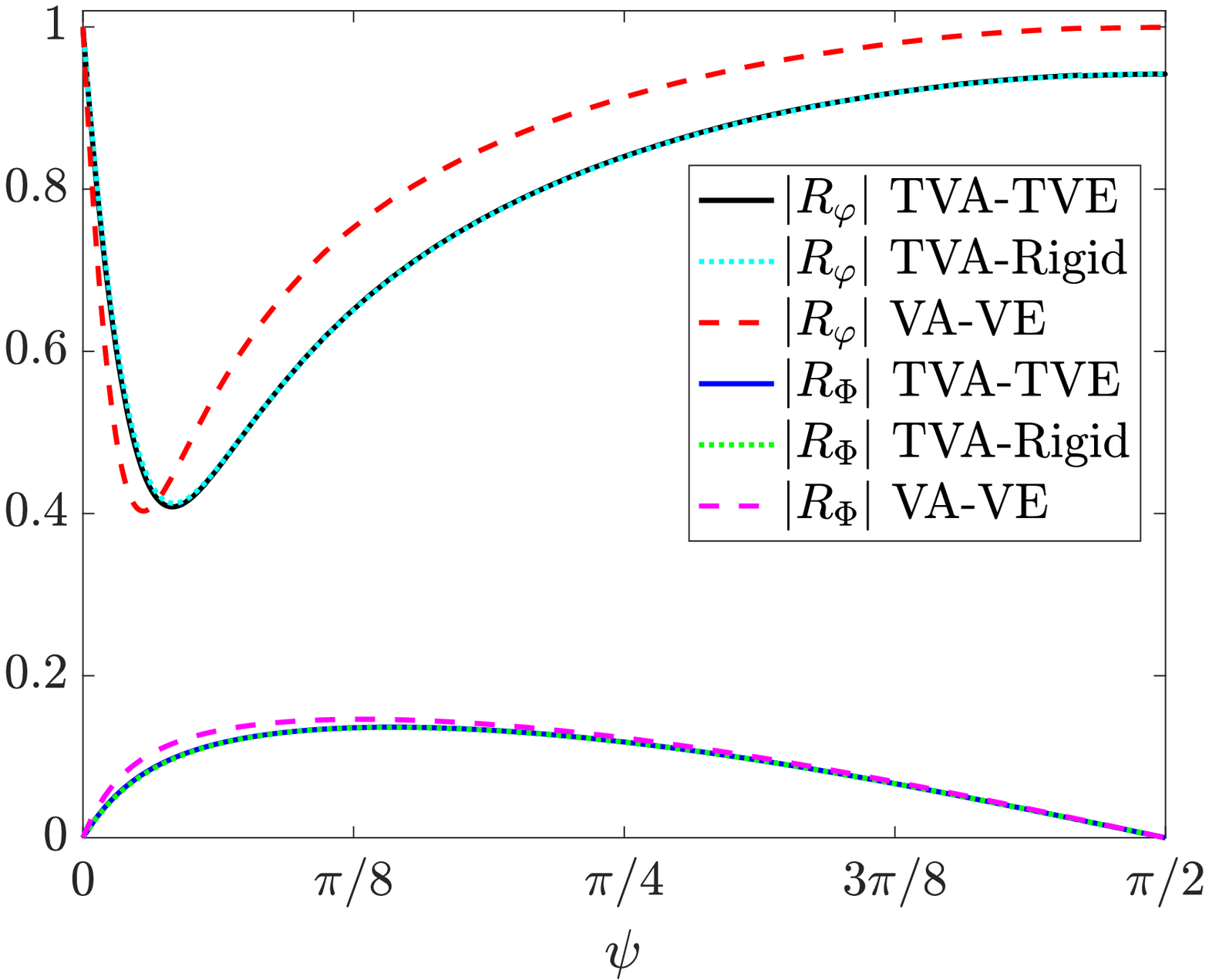}
    \caption{Air-Rubber 2 $f=10$ MHz.}
    \label{subfig:a)}
  \end{subfigure}
  \hfill
  \begin{subfigure}[t]{.45\textwidth}
    \centering
    \includegraphics[width=\linewidth]{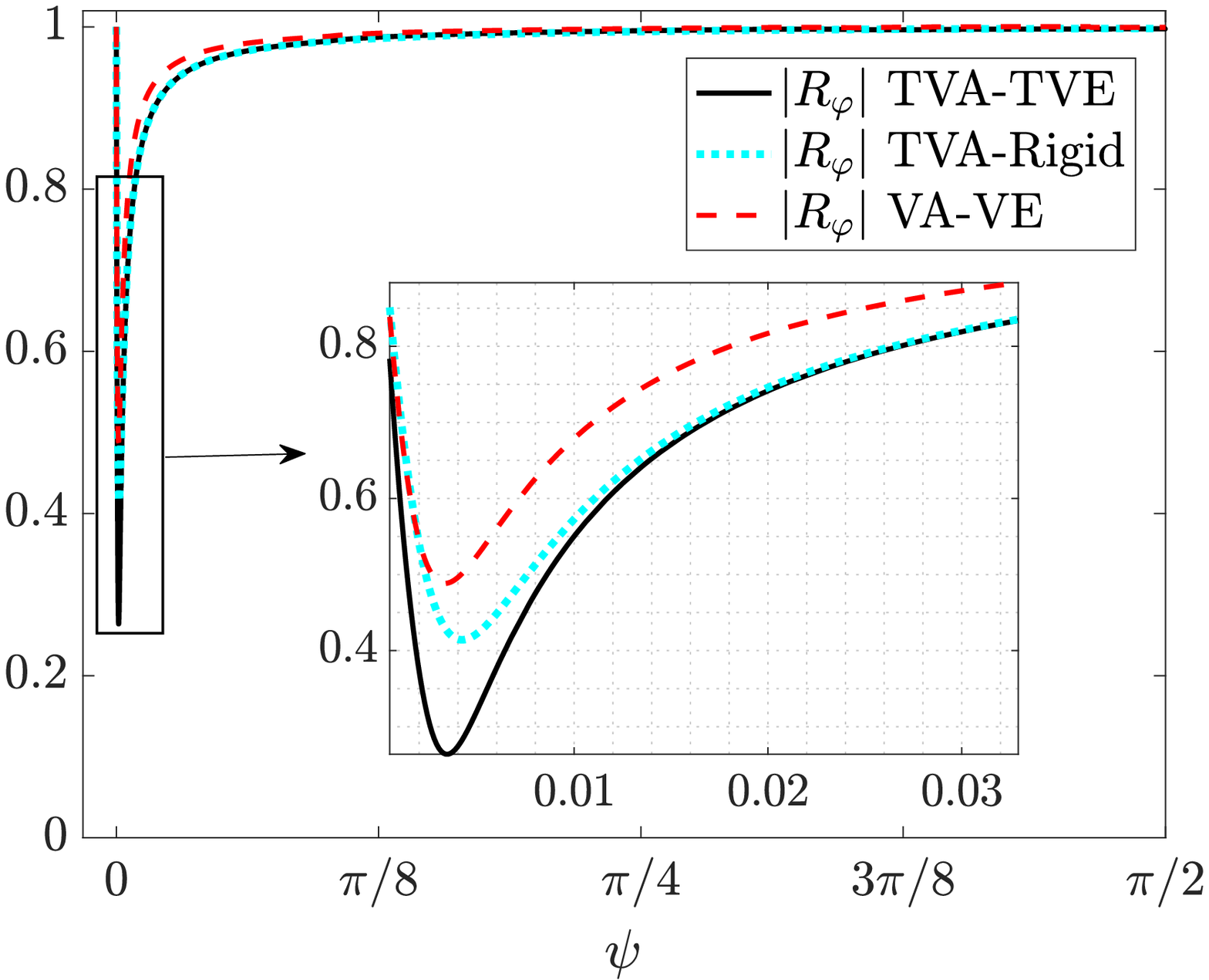}
    \caption{Air-Rubber 2 $f=10$ kHz}
     \label{subfig:b)}
  \end{subfigure}
  \smallskip
  \begin{subfigure}[t]{.45\textwidth}
    \centering
    \includegraphics[width=\linewidth]{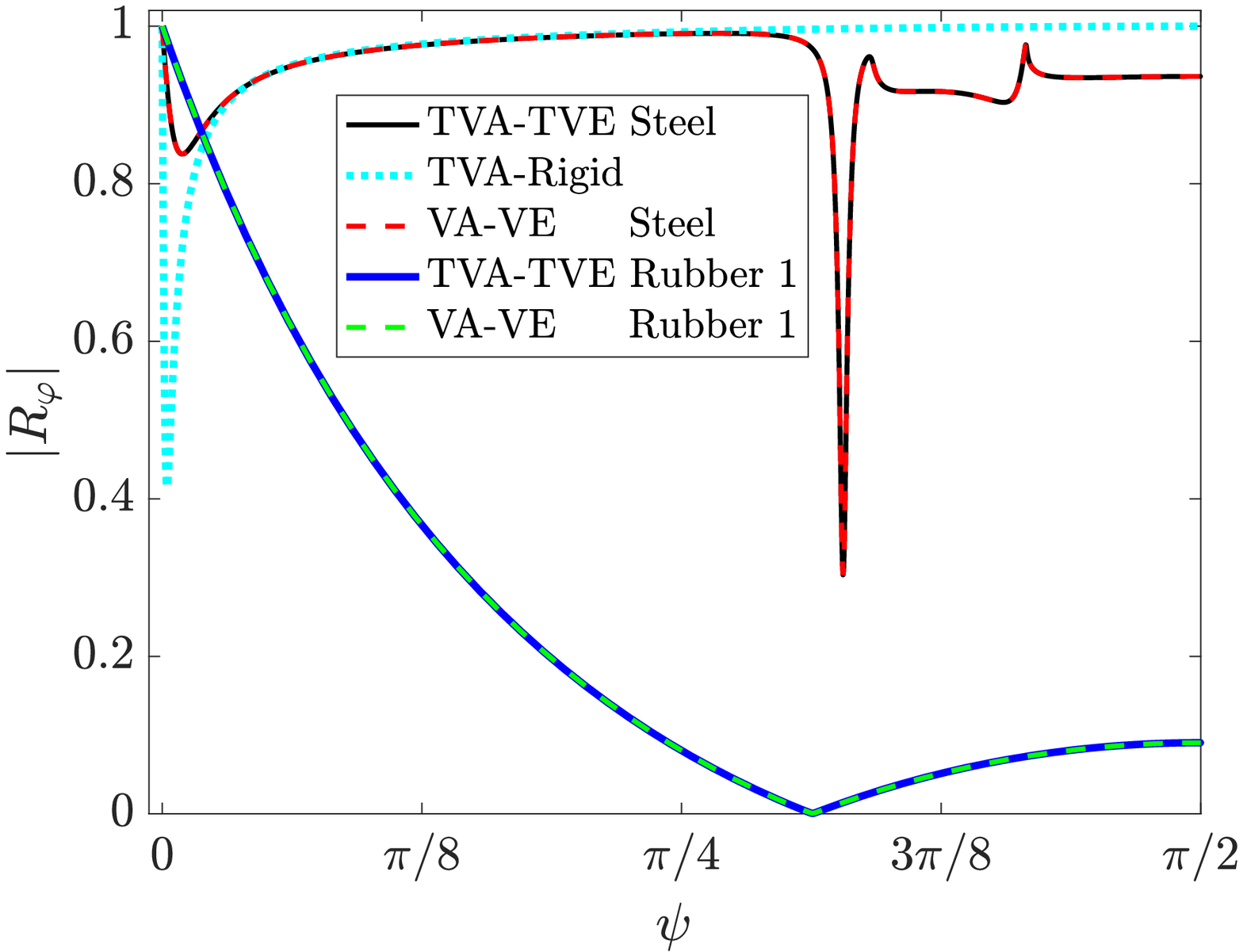}
    \caption{Water-Rubber 1/Steel $f=10$ MHz}
     \label{subfig:c)}
  \end{subfigure}
  \hfill
  \begin{subfigure}[t]{.45\textwidth}
    \centering
    \includegraphics[width=\linewidth]{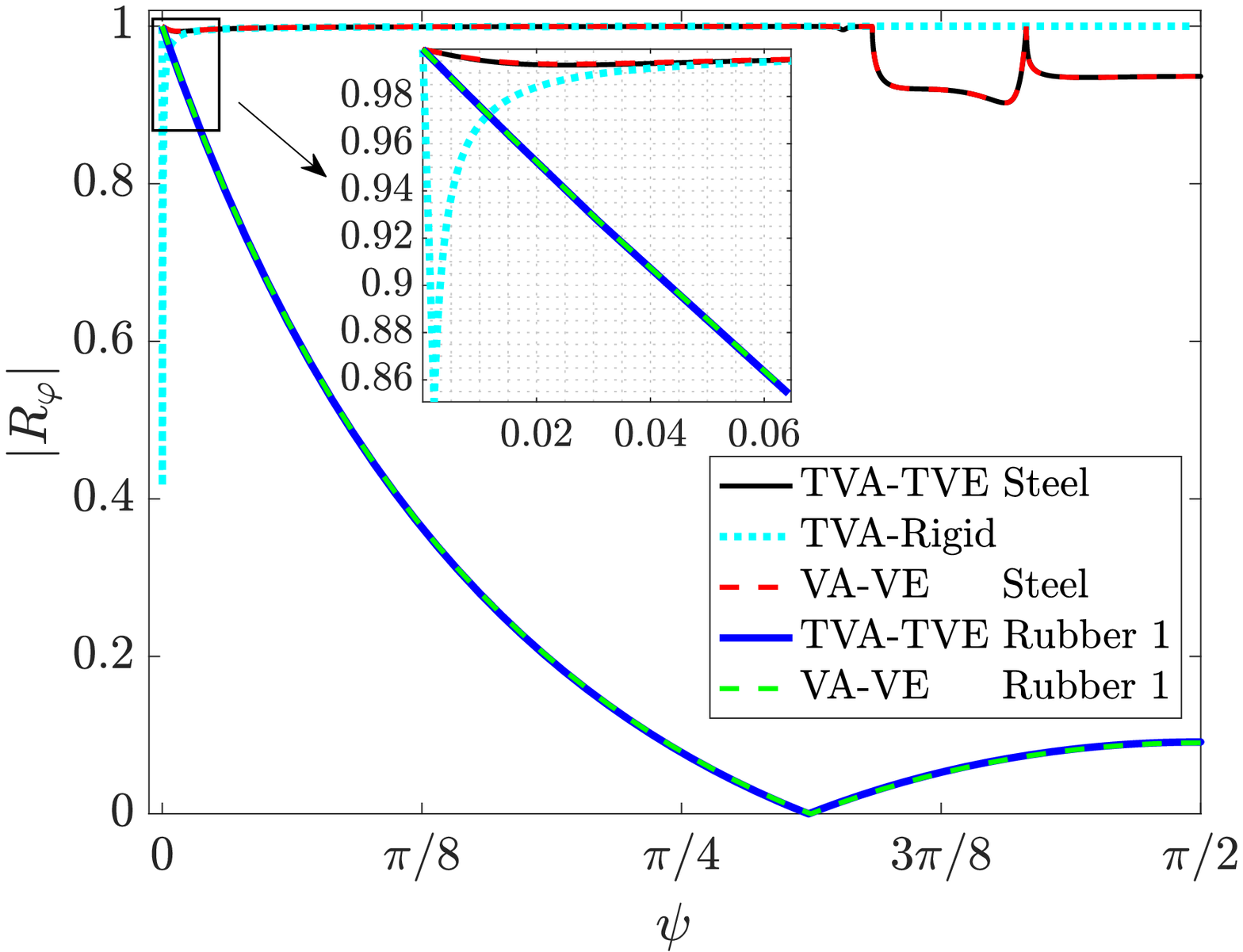}
    \caption{Water-Rubber 1/Steel $f=10$ kHz}
     \label{subfig:d)}
  \end{subfigure}
  \caption{Magnitude of the reflection coefficients as predicted by the different systems in Table \ref{table: Acronym} with the material constants used shown in Table~\ref{table: TVE parameters Rubber & Steel}. The results cover different fluid-solid interfaces for both higher ($f=10$MHz)  and lower ($f=10$kHz) frequencies, and the $x$-axis shows the angle of incidence $\psi$.}
\label{fig:Air-Rubber C_varphi, C_Phi}
\vspace{-.9cm}
\end{figure}

\noindent \textbf{Air-Solid interface.} Thermal effects are known to be important in air, as we can clearly see in Figure \ref{subfig:a)}, where various reflection coefficients are compared. This is evidenced by the value of thermo-mechanical coupling term for air (second of \ref{DIMTVE Lphi limit}) given by $|L_\theta| \approx 1$ due to air's high thermal expansion coefficient. The pressure dominated reflection coefficient $R_{\varphi}$ (responsible for most of the energy) is clearly different for a system which does not include thermal effects in air, such as VA-VE. This is especially true at higher frequencies, in agreement with \cite{cotterill2018thermo} for narrow slits. Here thermal effects for air are less pronounced for lower frequencies, as shown in Figure \ref{subfig:b)} for $f=10$ kHz. The reflected shear wave is no longer excited and $|R_\varphi|$'s minimum moves closer to the  grazing angle of incidence $\psi = 0$. This behaviour is due to viscous and thermal boundary layer effects near the interface, and can be described through an analytical expression for the \textit{specific admittance}, where the influence of frequency and angle of incidence become apparent, see e.g.\ Section 3.2.1 in \cite{bruneau2013fundamentals}. Naturally, the solution to the A-Rigid configuration in the absence of any losses gives $R_\varphi = 1$ everywhere, independently of the incident frequency, see (\ref{A1vsRigid SolutionsISO}).

Note that neither thermal nor viscous effects are important in medium 2, as using the rigid boundary conditions, TVA-rigid, accurately recovers the reflection coefficient of TVA-TVE. For all of these parameters we obtained almost identical results when swapping rubber for steel, noting that for air-steel the small discrepancy between TVA-rigid and TVA-TVE observed near grazing in  Figure \ref{subfig:b)} disappears. The overall excellent agreement is because in both cases there is little transmission into the solid. The same cannot be said of a water-solid interface as we now describe.\\

\noindent \textbf{Water-Solid interface.} 
As the mechanical impedance of water is closer to the impedance of most solids, more mechanical energy will be transmitted into the solid giving rise to fluid-strucutre interaction (FSI) effects.
This is apparent from Figures \ref{subfig:c)}, \ref{subfig:d)} where the TVA-rigid solutions no longer agree with the TVA-TVE system. On the other hand, in contrast to air, thermal effects are no longer particularly important, indicated by the fact that TVA-TVE and VA-VE solutions are almost the same. This is due to the smaller thermal coupling for water $|L_\theta| \approx 0.078$.
We observe that the $|R_\varphi|$ behaviour for water-Rubber 2 is indistinguishable at low and high frequencies, resembling the purely elastic solutions ((\ref{VE1vsVE2 Scat system}) with $\eta_{\mu_1}, \eta_{\mu_2}, \eta_{\lambda_1}, \eta_{\lambda_2}=0$) which are independent of frequency. The same can be said for the transmitted modes. Nevertheless, we will observe shortly how this behaviour can change when stress relaxation is considered.

For water-steel the frequency dependence is nevertheless apparent. In the TVA-TVE solutions, boundary layer effects are visible near grazing incidence at higher frequencies (Figure \ref{subfig:c)}), in contrast to the lower frequency regime, where $|R_\varphi|$ remains very close to one as seen in Figure \ref{subfig:d)}. The TVA-Rigid solutions greatly overestimate these effects near $\psi=0$ at both frequencies. As opposed to the in-air case, reflected boundary layer shear waves into the water were not found i.e. $|R_\Phi|\approx 0$ in each case and hence not included in the figures. The other notable frequency-dependent feature for water-steel is the emergence of a significant reduction in amplitude at high frequencies for a narrow range of angles of incidence around the interval ($\pi/4$, $3\pi/8$). This phenomenon was first observed experimentally in the 1960s for water-aluminum and it was noticed that it disagreed with predictions of elastic reflection–refraction theory. It has been discussed by several authors since including \cite{becker1970ultrasonic,deschamps1989liquid,borcherdt2009viscoelastic} where the latter reference provides a detailed explanation under a VA-VE model. Under the framework presented in this work, we have extended their model to include thermal losses in both media, although as we can observe these are not manifested in the solutions when compared to the isothermal solution. Finally, we note that the $|R_\varphi|$ behaviour for $\psi \in (3\pi/8,\pi/2)$ in Figures \ref{subfig:c)} and \ref{subfig:d)} is elastic and independent of frequency, and the two distinct features in this region are a consequence of the transmitted SV and P waves in the lower half-space being induced respectively (not shown).
\vspace{-0.3cm}

\subsubsection{Influence of stress relaxation}
We now explore the effect of stress relaxation in the solid. Little discussion is found on stress relaxation times for metals in the literature since in most instances they are nearly undamped materials e.g. \cite{liao2006estimation}, so here we focus on results for rubbery media following the values in Table \ref{table: TVE parameters Rubber & Steel}. \\

\noindent \textbf{Rubber-Air interface.} We first investigate a TVE-TVA interface, where the incident energy arises from the solid. Following the discussion in Section \ref{sec:linearTVE}\ref{subsection: general TVE}\ref{subsection:Choice of moduli}, we assume that the relaxation is purely in shear and is governed by a single-term Prony series, with the bulk modulus being a real valued constant such that
\begin{subequations}\label{moduli TVE-TVA}
\begin{align}
     \tilde{\mu}_1(\mi \omega) &= \mu_{\infty_1} - (\mu_{0_1} - \mu_{\infty_1}) \frac{\mi \omega t_r }{1 - \mi \omega t_r}, & \tilde{\lambda}_1(\mi \omega) &= K_1 -\frac{2}{3}\tilde{\mu}_1(\mi \omega), \\
     \tilde{\mu}_2(\mi \omega) &= - \mi \omega \eta_{\mu_2}, & \quad \tilde{\lambda}_2(\mi \omega) &= \lambda_2 - \mi \omega \eta_{\lambda_2}.
 \end{align}
\end{subequations}
As discussed in Section \ref{sec:linearTVE}\ref{subsection: general TVE}\ref{subsection:Choice of moduli}, the relevant non-dimensional parameter to investigate the different regions of the modulus is $\omega t_r$. For a given material, the relaxation time is fixed and it scales the resulting frequency behaviour. Here we choose three distinct values, namely $\omega t_r = 0.063, 1.005, 62.83$ corresponding to the rubbery, transition and glassy regions of the shear modulus, as shown explicitly in Table \ref{shear and poiss ratio rubbers}. 

\begin{figure}
\vspace{-0.5cm}
\centering
\includegraphics[width=.6\linewidth]{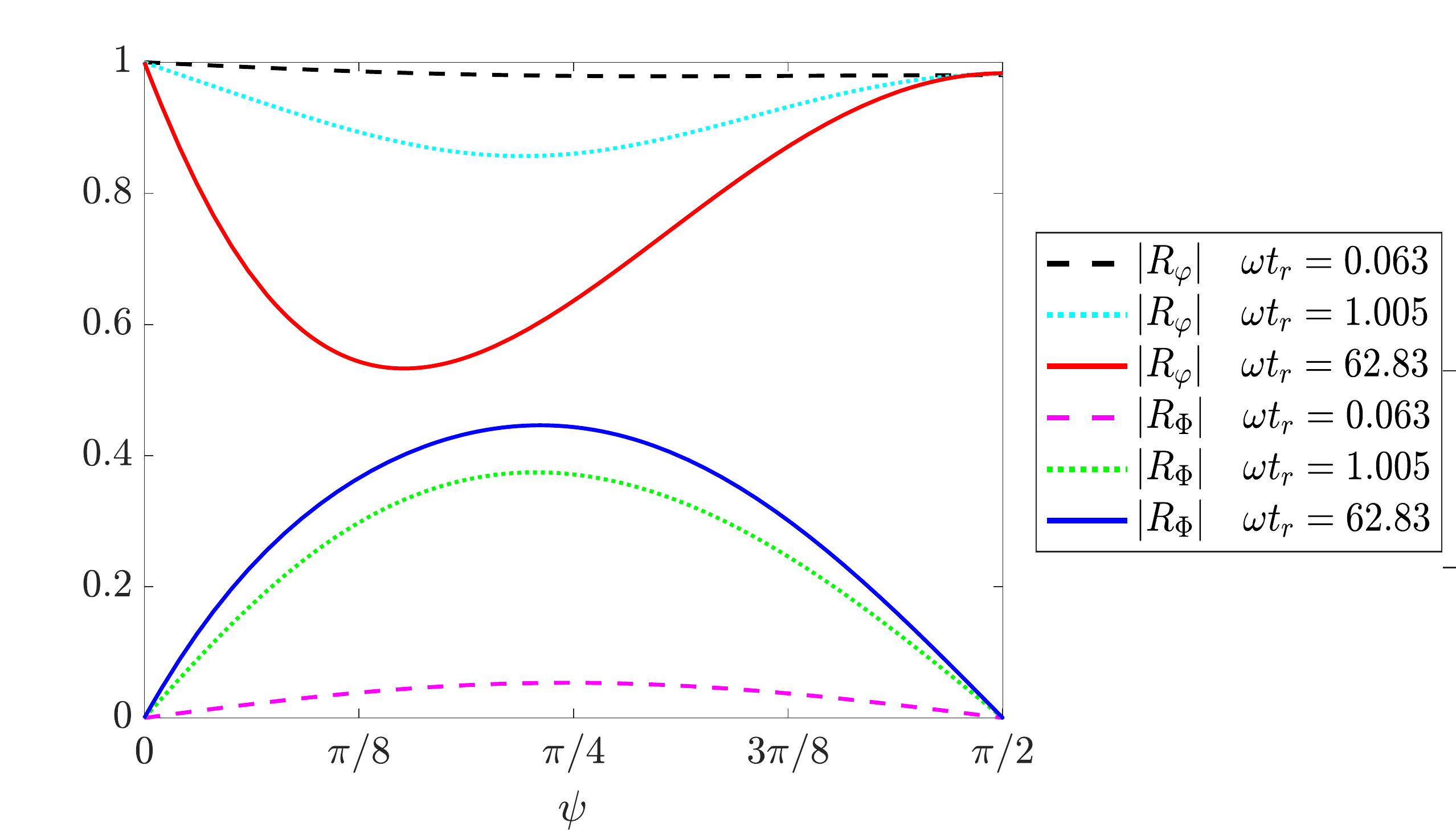}
\caption{Reflection coefficients for a Rubber2-air interface in the rubbery, transition and glassy regions of the shear modulus according to the SLSM. The material constants used are shown in Table~\ref{table: TVE parameters Rubber & Steel}.}
\label{Rubber 2-air relaxation}
\vspace{-0.7cm}
\end{figure}

For Rubber 1 in Table \ref{table: TVE parameters Rubber & Steel}, it was found that $R_\varphi \approx -1$, independently of $\psi, \omega t_r$. This is due to the fact that for Rubber 1 $ K_1 \gg |\Tilde{\mu}_1(\mi \omega)|$ at all frequencies since this material is nearly incompressible, and hence the associated Poisson's ratio remains very close to $1/2$ in each case. Nevertheless, for Rubber 2 the situation is much different, as shown in Figure \ref{Rubber 2-air relaxation}. In the rubbery region $\omega t_r = 0.063$, the incident angle dependence on reflection remains small but this changes in the transition region and especially in the glassy region. For $\omega t_r = 62.83$ we observe that the reflected SV wave gets excited with a global maximum near $\psi = \pi/4$ where the amplitude $R_\Phi$ becomes almost $50 \%$ of that of the incident wave. Despite the smaller ratio $\mu_{01}/\mu_{\infty_1}$ of Rubber 2 compared to Rubber 1, its higher magnitude implies that it becomes more compressible and the Poisson's ratio reduces (see Table \ref{shear and poiss ratio rubbers}) which in turn excites the reflected shear wave, e.g. for $\omega t_r = 1.005$, we have $|\Tilde{\nu}(\mi \omega)| =0.425$. Since these solutions are mainly influenced by the Poisson's ratio, for a practical realization it is the frequency dependence $\Tilde{\nu}(\mi \omega)$ that should be studied more in depth, see e.g. \cite{tschoegl2002poisson} for an extensive review.

\begin{table}
\centering
\small
\begin{tabular}{|c||c|c|}\hline
\backslashbox{$\omega t_r$}{Modulus}
&{${\Tilde{\mu}}$ (MPa) \quad  Rubber 1/Rubber 2}&{$\Tilde{\nu}$ \quad Rubber 1/Rubber 2 }
\\\hline\hline  
&&\\[-1em]
$0.062$ & $ 0.338 - 0.607 \mi $/$21 -  17 \mi$ & $0.4999 + 0.00018 \mi$/$0.489 + 0.008 \mi$ \\ \hline 
&&\\[-1em]
$1.005$ & $5.2 - 4.85  \mi$/$ 160 - 140 \mi$ & $0.498 + 0.00142 \mi$/$0.42 + 0.06 \mi$ \\ \hline 
&&\\[-1em]
$62.83$ & $9.99 - 0.154  \mi$/$299 - 4.4 \mi$ & $0.497 + 0.00004 \mi$/$0.36 + 0.0018 \mi$ \\ \hline 
\end{tabular}
\caption{Shear modulus and Poisson's ratio values according to the SLSM for various values of $\omega t_r$ covering the rubbery, transition and glassy regions of the two types of rubbers in consideration.}
\label{shear and poiss ratio rubbers}
\vspace{-1cm}
\end{table}

For both rubbers the thermo-mechanical coupling is small such that $L_\theta=O(10^{-2})$, and therefore equivalent results are obtained when using the VE-VE system (\ref{VE1vsVE2 Scat system}). Again due to the mechanical impedance mismatch, transmission into the air is negligible. In fact, these results obtained for air in the lower medium had excellent agreement with the associated problem of a single TVE half-space with traction free and isothermal/adiabatic boundary conditions. Although not included in this report, these simpler solutions showcase explicitly the role of $\nu$ described above (see e.g. \S 5.6 in  \cite{achenbach2012wave} in the absence of losses).\\

\noindent \textbf{Fluid-Rubber interface.} In the second example, we want to investigate whether stress relaxation effects in rubber can still alter the reflection/transmission pattern when the incident energy comes from the fluid, so we return to a fluid-solid TVA-TVE interface such that
\begin{subequations}\label{moduli TVA-TVE}
\begin{align}
    &\tilde{\mu}_1(\mi \omega) =  - \mi \omega \eta_{\mu_1},  &&\tilde{\lambda}_1(\mi \omega) = \lambda_1 - \mi \omega \eta_{\lambda_1},\\
    &\tilde{\mu}_2(\mi \omega) = \mu_{\infty_2} - (\mu_{0_2} - \mu_{\infty_2}) \frac{\mi \omega t_r }{1 - \mi \omega t_r}, && \tilde{\lambda}_2(\mi \omega) = K_2 -\frac{2}{3}\tilde{\mu}_2(\mi \omega).
\end{align}
\end{subequations}
In the case of air-rubber (1 \& 2), for each value of $\omega t_r$ the reflected modes behave as discussed with the local TVE model in Figures \ref{subfig:a)}, \ref{subfig:b)} and the transmission into the rubber is negligible. Although as we observed in Figure \ref{subfig:c)}, energy gets transmitted into the solid in a water-Rubber 1 interface, the frequency variation of the shear modulus according to the SLSM did not manifest in any results that deviated much from the Local TVE case. This occurs due to the high Poisson's ratio of Rubber 1, as discussed above for the rubber-air interface. For water-Rubber 2 however, significant differences in $|R_\varphi|, |T_\varphi|, |T_\Phi|$ do arise. \\

\begin{figure}
\centering
\begin{subfigure}{.5\textwidth}
  \centering
  \includegraphics[width=1\linewidth]{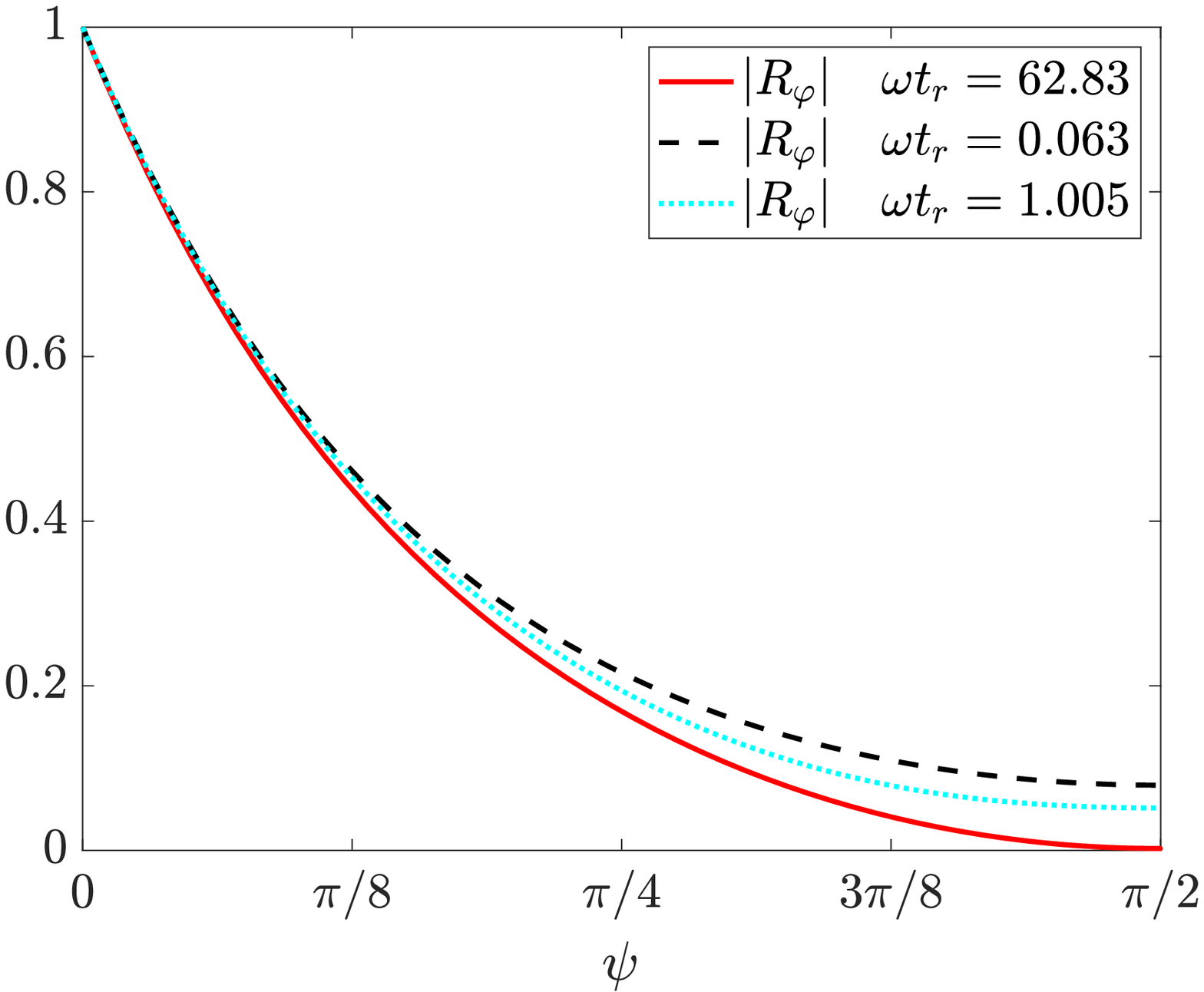}
\end{subfigure}%
\begin{subfigure}{.5\textwidth}
  \centering
  \includegraphics[width=1\linewidth]{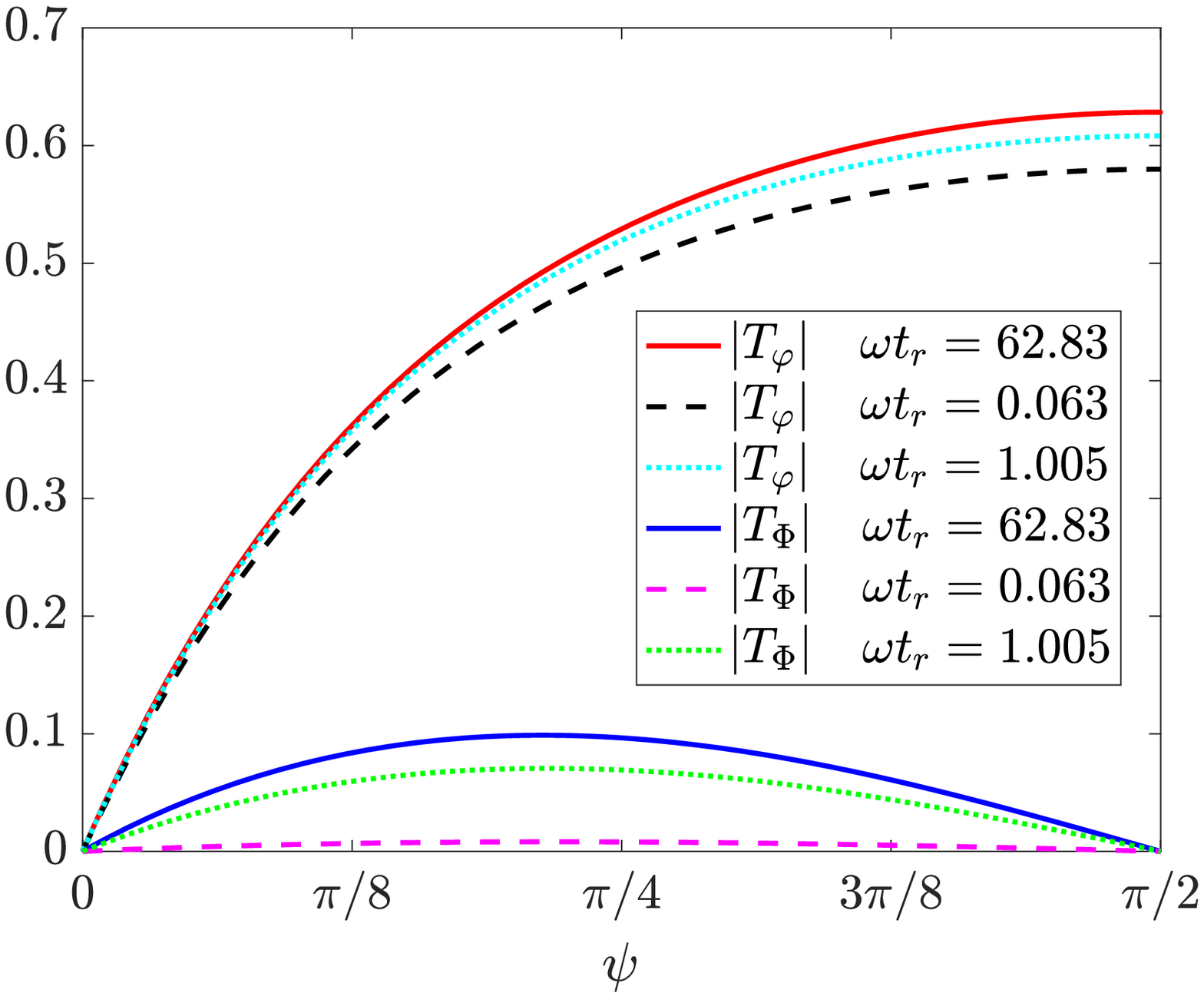}
\end{subfigure}
\caption{Reflection/Transmission of a water-Rubber 2 interface according to (\ref{TVE1vsTVE2 Scat system}), where the density of material 2 has been adapted to impedance match with $\mu_0$ in the glassy region.}
\label{water-rubber2 impedance matched}
\vspace{-0.6cm}
\end{figure}

It is often of interest in application to avoid any acoustic reflection in the incident medium, which requires impedance matching with the neighbouring medium. Since for these materials thermal coupling was found to be unimportant, (\ref{VE normal incident solutions}) can be used to tune Rubber 2 in order to impedance match it with the water for a particular value of frequency. As an illustration, following this principle we simply tune the density of Rubber 2 ($\Bar{\rho}_2: 2300 \rightarrow 1588$ kg$/$m$^3$) in order to impedance match it with water in the glassy region represented by $\omega t_r = 62.83$, as shown in Figure \ref{water-rubber2 impedance matched}. For the reflected/transmitted P waves, the differences between $\omega t_r$ increase monotonically as $\psi$ moves from grazing to normal incidence, where the maximum difference occurs. A $10 \%$ variation in the magnitude of the reflected amplitude was found between the glassy and rubbery regions. Similar values for this variation yield for the transmitted shear wave, where the maximum difference occurs near $\psi = \pi/4$.

\begin{table}
\centering
\footnotesize
\begin{tabular}{ 
 |p{4.9cm}||c|c|c|c|}
 \hline
 \multicolumn{5}{|c|}{ {\textbf{TVE Parameter Values}}} \\
 \hline
  Parameter (Symbol) [Unit] & Air & Water & Steel & Rubber 1 \& 2 \\
 \hline
  \textbf{Elastic}  & & & &\\ 
 Background density (${\rho}_0$) [kg m$^{-3}$] & 1.19 & 1000 &7932 & 1522 \& 2300\\ 
 Isothermal Bulk modulus  (K)  [Pa]  & 100.72$\times 10^{3}$ & 2.2$\times 10^{9}$ & 1.57$\times 10^{11}$ &  1.7$\times 10^9$ \& $ 10^9$\\ 
 Shear modulus  (${\mu}_0$) [Pa]  & 0 & 0 & 7.83$\times 10^{10}$ & $10^7$ \& 3$\times10^8$\\ 
 Relaxed Shear modulus for SLSM  (${\mu}_{\infty}$)  [Pa]  & - & - & - & 3$\times 10^5$\&2$\times 10^7$\\ 
  \hline \textbf{Local Viscous}   & & & &\\ 
 Dynamic shear viscosity (${\eta}_\mu$) [Pa s] & 1.8$\times 10^{-5}$ & $ 10^{-3}$ & 15 & $10^{-2}$\\ 
 Dynamic bulk viscosity (${\eta}_K$) [Pa s] & 1.1$\times 10^{-5}$ & 3$\times 10^{-3}$ & $10^{-8}$ &  $10^{-2}$\\ 
  \hline \textbf{Thermal}  & & & &\\ 
 Thermal conductivity (${\mathcal{K}}$)  [W m$^{-1}$ K$^{-1}$] &  0.026 & 0.597 & 30 & 2\\ 
 Specific heat at constant pressure (${c}_p$)  [J kg$^{-1}$ K$^{-1}$]  & 1005 & 4181.6 & 500 & 1300\\ 
 Ambient temperature (${T}_0$)  [K]  & 300 & 300 & 300 & 300 \\ 
 Coefficient of thermal expansion (${\alpha}$)  [K$^{-1}$]  & 1/300 & 2.6$\times 10^{-4}$  & 1.7$\times 10^{-5}$& 2.5$\times 10^{-4}$ \\ 
 Ratio of specific heats ($\gamma$) [-]   & 1.39 & 1.01  & 1.0003 & 1.008 \\
 \hline
\end{tabular}
\caption{Thermo-viscous parameter values for air, water, steel and rubber employed in the several plots of Section \ref{sec:Application1:2HSs}. Air is taken from \cite{pierce1981acoustics} and Water from engineeringtoolbox. The VE values for Steel are taken from Table 6.2.2. in \cite{borcherdt2009viscoelastic}, which follow from experiments. The high value of $\eta_\mu$ arises from `fitting' a Kelvin-Voigt model to the imaginary part of the shear modulus which comes from measurements at 10MHz. The values of rubber are based on the ranges provided in \cite{AzoRubber}.}
\label{table: TVE parameters Rubber & Steel}
\vspace{-0.9cm}
\end{table}

\vspace{-0.5cm}
\section{Conclusions}\label{sec:conclusions}
Understanding how to model and exploit loss mechanisms in complex materials is important in many applications and increasingly so in the areas of composite media and metamaterials science. Here we have presented a general unified framework, permitting the incorporation of both creep and relaxation via time non-locality, with which one can study linear wave propagation in thermo-visco-elastic media. We illustrated the framework with the configuration of two semi-infinite half-spaces in perfect contact, with plane compressional-wave incidence on the interface that separates the media. We used this example to compare solutions when incorporating viscosity and thermal effects. For fluid-solid interfaces we noted the important role of the incident frequency and angle on the contribution to visco-thermal effects as well as visco-elastic attenuation within the solid. For the latter we emphasized the differences induced when the shear modulus includes stress relaxation, as opposed to the local-in-time counterpart where the real part of the modulus remains fixed.

There are many advantages to the unified framework presented here, but three are key. Firstly, it provides a mechanism to study canonical wave propagation problems when there is coupling between different media, and specifically between what are classically perceived as \textit{fluid} and \textit{solid}. As we have shown this distinction is often clear away from boundaries but is less clear close to such interfaces. A unified framework allows such modelling to be carried out once and for all, without the need to develop separate models for each, as is often done~\cite{karlsen2015forces, cotterill2018thermo}. To help illustrate the connection between the framework and other well-known models for dissipation, such as thermo-visco-acoustics in fluids, or visco-elasticity, we have demonstrated how to take the appropriate limits to recover these special cases from our framework.

The second key advantage is the potential use of the framework to understand fully time-dependent problems. It is common for wave propagation problems to be studied at single frequencies, which is sufficient in its own right, but if a viscoelastic model is employed, one must be confident that this model is capable of representing the behaviour across a broad range of frequencies, especially if one wishes to subsequently use this model in the time domain, given that a time domain signal will encompass a vast range of frequency content in general. It is often seen as standard practice to employ simple Kelvin-Voigt models to account for visco-elasticity, with "parameters that are frequency dependent'' \cite{favretto1997excitation,favretto1996theoretical}. Whilst this may be sufficient to model the material response at fixed frequencies, it is not sufficient to be employed in the time domain.

The third advantage of the unified framework that incorporates stress relaxation and creep compliance is that one can then employ these models to understand and describe wave propagation in polymeric media. Such materials have the behaviour as illustrated throughout this paper, with a specific frequency at which maximum loss occurs, also related to a temperature, know as the glass-transition. This behaviour is particularly important to accommodate when polymers are employed in the metamaterial context since the design of metamaterials focuses on internal resonance and therefore one may wish to design this resonance with knowledge of this transition in mind, either to increase or decrease inherent attenuation in the material.

We anticipate that the presented framework can now be employed on various problems of interest. In particular it can be used to unify the approach to the problem described in \cite{cotterill2018thermo} and this will be extended in upcoming work.



\dataccess{We provide the code to generate all the graphs in \cite{artGitHub}.}

\aucontribute{All authors conceived the idea and design of the study. E.G.N.\ set out the theoretical
model based on previous theoretical developments formulated by A.G., V.P and W.J.P. E.G.N. extended this to incorporate stress relaxation effects and implemented the model for the exams shown, given extensive discussions with R.A., D.N.\ and W.J.P. Limits were studied by A.G. and E.G.N. E.G.N. wrote the initial draft of the paper and this was edited by, and iterated with, all other authors. All authors gave final
approval for publication.}

\competing{The authors have no competing interests.}

\funding{E.G.N. is grateful to Thales UK and the Engineering and Physical Sciences Research Council (EPSRC, UK) for his PhD CASE studentship. W.J.P. is grateful to EPSRC for funding his Fellowship (EP/L018039/1) and Fellowship extension (EP/S019804/1). A.G., V.P. and W.J.P. also acknowledge EPSRC funding via grant EP/M026205/1.}

\appendix
\section{Local isotropic TVE stress-strain and entropy relations}\label{local TVE derivation}

By assuming that the free energy $\Psi$ can be written as a function of the strain $\bm \varepsilon$ and temperature $T$ and given its relationship with the Cauchy stress (\ref{C-Stress: TE+VE}), the strain energy can be written explicitly as a series expansion about $\bm \varepsilon = \bm 0$ and $T = {T}_0$, up to second order in $\bm \varepsilon$, and ${T}$. This leads to 
\begin{align*}\label{FreeEnergy1}
    {\rho}_0 {\Psi}(\bm \varepsilon, {T}) ={\rho}_0\Bigg( \left.{\Psi}\right\rvert_{\bm  \varepsilon=\bm 0,{T}={T}_0}+ 
    &
    \left.\frac{\partial {\Psi}}{\partial \bm \varepsilon}\right\rvert_{\bm 0,{T}_0} \hspace{-0.35cm} \bm :  \bm \varepsilon 
    + \left.\frac{\partial {\Psi}}{\partial {T}}\right\rvert_{\bm 0,{T}_0} \hspace{-0.35cm} ({T}-{T}_0)
    + \frac{1}{2!} \bm \varepsilon \bm : \left.\frac{\partial^2 {\Psi}}{\partial \bm \varepsilon \partial \bm \varepsilon}\right\rvert_{\bm 0,{T}_0} \hspace{-0.35cm} \bm : \bm \varepsilon  \numberthis \\
    &+\left.\left.\frac{\partial^2 {\Psi}}{\partial {T}^2}\right\rvert_{\bm 0,{T}_0} \hspace{-0.35cm}\frac{({T}-{T}_0)^2}{2!}
     + 2 \frac{({T}-{T}_0)}{2!}  \left.\frac{\partial^2{\Psi}}{\partial \bm \varepsilon\partial {T}}\right\rvert_{\bm 0,{T}_0} \hspace{-0.35cm} \bm : \bm \varepsilon \right),
\end{align*}
where we have assumed that $\bm \varepsilon, ({T}-{T}_0)/{T}_0  \ll 1$ and both are of the same order. If we further assume isotropy, we can reach
\begin{subequations}\label{isotropy local TVE}
\begin{align}\label{eqn:Tensor1}
    & {\rho}_0\left.\frac{\partial {\Psi}}{\partial \bm \varepsilon}\right\rvert_{\bm 0,{T}_0}  \bm : \bm \varepsilon = {a}_0\tr (\bm \varepsilon), 
      &&\left.\frac{\partial^2 {\Psi}}{\partial {T}^2}\right\rvert_{\bm 0,{T}_0} = -\frac{{c}_v}{ {T}_0},
    \\
    & {\rho}_0 \bm \varepsilon : \left.\frac{\partial^2 {\Psi}}{\partial \bm \varepsilon \partial \bm \varepsilon}\right\rvert_{\bm 0,{T}_0} \hspace{-0.3cm} \bm : \bm \varepsilon  = \lambda (\tr \bm \varepsilon)^2 + 2\mu \tr (\bm \varepsilon^2),   &&{\rho}_0 \left.\frac{\partial^2 {\Psi}}{\partial \bm \varepsilon \partial {T}}\right\rvert_{\bm 0,{T}_0} \hspace{-0.35cm} \bm : \bm \varepsilon = - {\alpha} K\tr (\bm \varepsilon),
\end{align}
\end{subequations}
where $K =  {\lambda} + 2{\mu /3}$ denotes the isothermal bulk modulus, and the material constants $\mu, \lambda,c_v,\alpha, a_0$ have conveniently been chosen to fit standard conventions. Using (\ref{eqn: TE thermodynamic restrictions}), and (\ref{isotropy local TVE}) we may rewrite (\ref{FreeEnergy1}) as
\begin{align*}\label{eqn:Psi_General1}
   {\rho}_0 {\Psi}(\bm \varepsilon, {T}) &= {\rho}_0({\mathcal{E}}_0-{T}_0 {\mathfrak{s}}_0-{\mathfrak{s}}_0({T}-{T}_0) - \frac{{c}_v}{ 2{T}_0}({T}-{T}_0)^2) + {a}_0 \tr \bm \varepsilon \numberthis\\
   &+\frac{1}{2}\left(\lambda (\tr \bm \varepsilon)^2 + 2\mu\tr (\bm \varepsilon^2) \right) - {\alpha}K ({T}-{T}_0) \tr(\bm \varepsilon).
\end{align*}
From the above, (\ref{eqn: TE thermodynamic restrictions}), and (\ref{C-Stress: TE+VE}) it follows that the Cauchy stress tensor and entropy become
\begin{subequations}
\begin{align}\label{stress a_i}
  & {\bm \sigma} = {a}_0\bm I + \lambda \tr(\bm \varepsilon)\bm I + 2 \mu \bm \varepsilon - \alpha K ({T}-{T}_0) \bm I + 2 {\eta}_\mu \dot {\bm \varepsilon} + \left({\eta}_K - \frac{2{\eta}_\mu}{3}\right) \bm I \tr(\dot  {\bm \varepsilon}), 
  \\ 
  \label{entropy a_i}
  & {\mathfrak{s}} = {\mathfrak{s}}_0 +\frac{{c}_v}{ {T}_0} ({T}-{T}_0) + \frac{\alpha K}{{\rho}_0} \tr \bm \varepsilon.
\end{align}
\end{subequations}
We can let ${a}_0=0$ since we are not considering any form of pre-stress. By comparing with (\ref{eqn:stress and entropy}), we can now identify: ${\lambda}$ and ${\mu}$ as the (isothermal) Lam\'{e} coefficients,
${c}_v={T}_0(\partial {\mathfrak{s}}/ \partial {T})_{\bm \varepsilon =\bm{0}}$ as the specific heat at constant deformation (see e.g. Article 1.12 in \cite{boley2012theory}), and ${\alpha}$ as the coefficient of volumetric thermal expansion\footnote{Note that for an isotropic material, this term is three times the coefficient of linear thermal expansion, which is also commonly found in the thermodynamic literature.} ${\alpha} = \left({\partial \tr({\bm \varepsilon})/\partial {T}} \right)_{\bm \varepsilon =\bm{0}}$. Equivalent theories for TVE can be derived similarly, in particular if (\ref{FreeEnergy Local theory}) is replaced with the Gibbs energy, the specific entropy can be written in terms of stress as (see e.g. \cite{lubarda2004thermodynamic} equation (34))
\begin{equation}
    \label{entropy wrt sigma Gibbs energy}
    {\mathfrak{s}} = {\mathfrak{s}}_0 + \frac{{\alpha}}{3 {\rho}_0}\tr \bm {\sigma} + {c}_p \theta,
\end{equation}
where similarly ${c}_p={T}_0(\partial {\mathfrak{s}}/ \partial {T})_{\bm {\sigma}=\bm{0}}$ is defined as the specific heat at constant deformation of the solid in consideration. When $\tr (\dot{\bm\varepsilon})$ can be neglected in (\ref{stress a_i}), then we can write (\ref{entropy a_i}) in terms of stress to obtain
\begin{equation}\label{entropy wrt sigma}
    {\mathfrak{s}} = {\mathfrak{s}}_0 + {c}_v \theta + \frac{{\alpha}}{3{\rho}_0}\Bigg(\tr \bm {\sigma} + 3 {\alpha} {K} ({T}-{T}_0) \Bigg),
\end{equation}
which can be directly equated with (\ref{entropy wrt sigma Gibbs energy}) in order to obtain the identity (\ref{eqn:thermodIdentity}).

\vspace{-0.2cm}
\section{Non-dimensionalisation and convenient physical limits}\label{appendix: section 4 details}
\subsection{Non-dimensionalisation}\label{appendix:non-dimensionalization}

For the numerical implementation it is convenient to re-write the dimensional equations with non-dimensional quantities. We choose to non-dimensionalise with respect to the thermo-elastic quantities from (the top) medium 1. In particular, we choose $\bar{c}_1$ to denote the (adiabatic) longitudinal speed of sound of the upper material in the lossless case, i.e. $\bar{c}_1^2=(\bar{\lambda}_1+2\bar{\mu}_1)/\bar{\rho}_1$ and $\bar{\ell}$ represents an arbitrary length scale. {In order to distinguish between dimensional/non-dimensional quantities here, we write all dimensional quantities with an overbar.}
\begin{align*}
    \nabla = \bar{\ell} \overline{\nabla}, \quad \omega &= \frac{\bar{\ell}}{\bar{c}_1} \bar{\omega}, \quad \{\mathbf{u}_m,\mathbf{x} \}= \frac{1}{\bar{\ell}}\{\bar{\mathbf{u}}_m,\bar{\mathbf{x}} \}, \quad \{\phi_m, \varphi_m, \vartheta_m, \Phi_m \}= \frac{1}{\bar{\ell^2}} \{\bar{\phi}_m, \bar{\varphi}_m, \bar{\vartheta}_m, \bar{\Phi}_m \},\\
    \mathscr{K}_m &= \frac{\bar{T}_1 \bar{\mathscr{K}}_m}{\bar{\rho}_1 \bar{c}_1^3 \bar{\ell}}, \quad c_{v_m} = \frac{\bar{T}_1}{\bar{c}_1^2} \bar{c}_{v_m}, \quad \alpha_m = \bar{\alpha}_m \bar{T}_1, \quad  \{\eta_{\lambda_m}, \eta_{\mu_m} \}= \frac{1}{\bar{\rho}_1 \bar{c}_1 \bar{\ell}} \{\bar{\eta}_{\lambda_m}, \bar{\eta}_{\mu_m}\},\\
    \{\tilde{\lambda}_m, \tilde{\mu}_m, \tilde{K}_m, \bm{\sigma}_m  \}&= \frac{1}{\bar{\rho}_1 \bar{c}_1^2} \{\bar{\tilde{\lambda}}_m, \bar{\tilde{\mu}}_m, \bar{\tilde{K}}_m, \bar{\bm{\sigma}}_m  \}, \quad \{k_{\theta_m}^2, k_{\phi_m}^2, k_{\Phi_m}^2 , L_{\phi_m}\} = \bar{\ell}^2 \{\bar{k}_{\theta_m}^2, \bar{k}_{\phi_m}^2, \bar{k}_{\Phi_m}^2 , \bar{L}_{\phi_m}\},\\
    k_{\theta_1}^2 = \frac{\mi \omega c_{v_1}}{\mathscr{K}_1}&, \quad k_{\phi_1}^2 = \frac{ \omega^2}{\tilde{\lambda}_1 + 2 \tilde{\mu}_1}, \quad k_{\Phi_1}^2 = \frac{\omega^2}{\tilde{\mu}_1}, \quad L_{\phi_1}=  \frac{\mi \alpha_1 \omega K_1}{\mathscr{K}_1}, \quad L_{\theta_1}= - \frac{\alpha_1 K_1}{\tilde{\lambda}_1 + 2\tilde{\mu}_1},\\
     k_{\theta_2}^2 =  \frac{\mi \omega c_{v_2}}{\mathscr{K}_2}r &, \quad k_{\phi_2}^2 = \frac{ \omega^2}{\tilde{\lambda}_2 + 2 \tilde{\mu}_2}r, \quad k_{\Phi_2}^2 = \frac{\omega^2}{\tilde{\mu}_2}r, \quad L_{\phi_2}= \frac{\mi \alpha_2 \omega K_2}{\mathscr{K}_2}, \quad L_{\theta_2}= - \frac{\alpha_2 K_2}{\tilde{\lambda}_2 + 2\tilde{\mu}_2}\left(\frac{\bar{T}_2}{\bar{T}_1} \right),
\end{align*}
where $m=1,2$ depending on the medium, $r=\bar{\rho}_2/\bar{\rho}_1$ is the contrast parameter and the background temperature ratio $\bar{T}_2/\bar{T}_1=1$ due to continuity of temperature across the boundary.

\subsection{TVE-TVE scattering system}
This matrix system and its derivation is provided in an open access Mathematica file in \cite{artGitHub}.

\vspace{-0.2cm}
\subsection{VE-VE scattering system}
In Section \ref{sec:theorylimits}\ref{subsection:TVE->VE} we learned how to recover the theory of isothermal visco-elasticity (VE) from that of TVE. For completeness purposes, we next formulate the scattering problem in Section \ref{sec:Application1:2HSs} for such media. This problem is well discussed in the VE literature, see e.g. \cite{borcherdt2009viscoelastic} Section 5.3. Equations (\ref{TVE-TVE incident P wave})-(\ref{TVE 2 HS BCs 1})
are replaced by
\begin{subequations}\label{VE HS R/T amplitudes}
\begin{align} 
\phi_\text{I} &= \me^{- \mi k_{\phi_1}(x \cos \psi + y \sin \psi)}, \qquad \psi \in[0, \pi],\\
\label{VE HS P wave reflected amplitudes}
\phi_\text{R} &= R_{\phi } \me^{ \mi k_{\phi_1 } \sin \psi y}\me^{- \mi k_{\phi_1}x \cos \psi},  
& \phi_\text{T} &= T_{\phi } \me^{- \mi d_{\Phi_\text{T}} y}\me^{- \mi k_{\phi_1}x \cos \psi},\\ \label{VE HS SV wave reflected amplitudes}
    \mathbf \Phi_\text{R} & = \mathbf{e_z} R_{\Phi} \me^{ \mi d_{\Phi_\text{R}} y} \me^{- \mi k_{\phi_1}x \cos \psi},  &
    \mathbf \Phi_\text{T} & = \mathbf{e_z} T_{\Phi} \me^{- \mi d_{\Phi_\text{T}} y}\me^{- \mi k_{\phi_1}x \cos \psi},
\end{align}
\end{subequations}
and $d_{\Phi_\text{T}} =  \mi \sqrt{-(k^2_{\Phi_2} - k^2_{\Phi_1} \cos^2 \psi)}$,  
    $d_{\Phi_\text{R}} = \mi \sqrt{-(k^2_{\Phi_1} - k^2_{\Phi_1} \cos^2 \psi)}$ and $d_{\Phi_\text{T}} = \mi \sqrt{-(k^2_{\Phi_2} - k^2_{\Phi_1} \cos^2 \psi)}$, ensures that $ \operatorname{Im} d_{\Phi_\text{T}}, \operatorname{Im}d_{\Phi_\text{R}},
    \operatorname{Im}d_{\Phi_\text{T}} 
    \geq 0$. The BCs reduce to $\hat{\mathbf{u}}^{\text{VE}}_1 = \hat{\mathbf{u}}^{\text{VE}}_2$ and $\hat{\boldsymbol \sigma}^{\text{VE}}_1  \mathbf{e_y} = \hat{\boldsymbol \sigma}^{\text{VE}}_2 \mathbf{e_y}$
which must hold on $y=0$. The application of these 4 BCs will determine the unique four constants $\{R_\phi, T_\phi, R_\Phi, T_\Phi\}$, see equation (5.3.21) in \cite{borcherdt2009viscoelastic} or \cite{artGitHub} for explicit details.

\vspace{-0.2cm}
\subsection{TVA-Rigid scattering problem}
In Section \ref{sec:theorylimits}\ref{subsection: TVE-> TVA} we discussed how local TVE yields the classical TVA theory for fluids in the limit of vanishing shear modulus, so we let $\mu \rightarrow 0$. For a thermo-viscous fluid in contact with a rigid interface at $y=0$, we impose no-slip and for the temperature field an isothermal boundary condition, that is $-\mi \omega \mathbf{u}^\text{TVA} = 0$ and $\theta^\text{TVA} = 0$
on $y=0$, noting that we have dropped the subscript since here we are only considering motion on $y\geq 0$. Following (\ref{TVE HS R/T amplitudes}) our fields are given by
\begin{subequations}\label{TVE Halfspace Fields ISO ALL}
\begin{align}\label{TVE Halfspace Fields ISO}
    \varphi &= \me^{-\mi k_{\varphi} x \cos \psi} \left( \me^{-\mi k_{\varphi } y \sin \psi} + R_\varphi \me^{\mi k_{\varphi }  y \sin \psi} \right),\\
    \vartheta &= R_\vartheta \me^{- \mi k_{\varphi }  x \cos \psi +\mi d_\vartheta y},  \qquad     \Phi = R_\Phi \me^{- \mi k_{\varphi }  x \cos \psi +\mi d_\Phi y},
\end{align}
\end{subequations}
with $d_\vartheta = \mi \sqrt{-(k_{\vartheta}^2 - k_{\varphi}^2 \cos^2 \psi)}$, $d_\Phi = \mi \sqrt{-(k_{\Phi}^2 - k_{\varphi}^2 \cos^2 \psi) }$ which ensures $\operatorname{Im} d_{\vartheta} , \operatorname{Im}d_{\Phi} \geq 0$. Substitution of (\ref{TVE Halfspace Fields ISO ALL}) into the governing equations (\ref{eqn:decoupledCURLpart}), (\ref{TVE LOcal 2 bulk PDEs}) and using (\ref{eqn:diplacementDecompos}), (\ref{eqn:coupledphi}) for the boundary conditions given above (\ref{TVE Halfspace Fields ISO ALL}), we obtain exact expressions
\begin{align}
\left(\begin{array}{c}
     R_\varphi  \\
     R_\vartheta \\
     R_\Phi
\end{array}\right)=
  \left(
\begin{array}{c}
 \frac{  -\cos^2 \psi +  \mathcal{B} (\mathcal{F}+\mathcal{G})}{ \cos^2 \psi - \mathcal{B} (\mathcal{F}-\mathcal{G})} \\
 \frac{  -2\sin \psi \mathcal{F}}{\cos^2 \psi- \mathcal{B} (\mathcal{F}-\mathcal{G})}\frac{k_\varphi}{d_\vartheta}\\
 \frac{ \sin 2\psi }{\cos^2 \psi- \mathcal{B} (\mathcal{F}-\mathcal{G})} \\
\end{array}
\right),  \label{TVE1vsRigid SolutionsISO}
\end{align}
where 
\begin{equation} \label{TVE1vsRigid Iso Definitions}
\mathcal{B}
= \sqrt{\frac{k_{\Phi }^2}{k_{\varphi }^2}-\cos^2 \psi}, \quad
    \mathcal{F} = \frac{d_\vartheta \mathscr{T}_\varphi}{k_\varphi(\mathscr{T}_\vartheta - \mathscr{T}_\varphi)}, \quad \mathcal{G}=\frac{\mathscr{T}_\vartheta \sin \psi}{\mathscr{T}_\vartheta - \mathscr{T}_\varphi}.
\end{equation}
With the current potentials (\ref{TVE Halfspace Fields ISO ALL}), the energy balance in this case reduces to
\begin{equation}\label{Single TVE HS: conservation of energy}
E_{R_\varphi}+E_{R_\vartheta} + E_{R_\Phi}+ E_{IR_\text{IR}}+ E_{IR_\text{RR}}=1,
\end{equation}
which are defined in electronic supplementary material Section SM4 and we must use the current potentials (\ref{TVE Halfspace Fields ISO ALL}). The visco-acoustic VA solution can be directly obtained from (\ref{TVE1vsRigid SolutionsISO}), (\ref{TVE1vsRigid Iso Definitions}) by letting $\mathscr{T}_\varphi \rightarrow 0$ which results in $\mathcal{F}\rightarrow 0$ and $\mathcal{G}\rightarrow \sin \psi$ so that (\ref{TVE1vsRigid SolutionsISO}) becomes
\begin{equation}\label{VE1vsRigid SolutionsISO}
     R_\varphi \rightarrow \frac{  -\cos^2 \psi +  \mathcal{B} \sin \psi}{ \cos^2 \psi + \mathcal{B}\sin \psi},  \qquad  R_\vartheta, \rightarrow 0, \qquad R_\Phi \rightarrow  \frac{ \sin 2\psi }{\cos^2 \psi+ \mathcal{B}\sin \psi}. 
\end{equation}
Finally, for the purely acoustic solution in the absence of any losses, we must further let $\eta_\mu \rightarrow 0$, which results in $\mathcal{B}\rightarrow \infty$, obtaining only the trivial solution
\begin{equation}\label{A1vsRigid SolutionsISO}
     R_\varphi \rightarrow 1, \qquad  R_\vartheta, R_\Phi \rightarrow 0.
\end{equation}








\section*{References}
\begingroup

\renewcommand{\section}[2]{}

\bibliographystyle{unsrtnat}
\bibliography{Bibliography.bib}
\endgroup

\end{document}